\documentclass[aps,prb,twocolumn]{revtex4}
\usepackage{amsfonts}
\usepackage{amsmath}
\usepackage{amssymb}
\usepackage[final]{graphicx}
\begin{document}
\title{Resonant electromagnetic emission from intrinsic Josephson-junction
stacks with laterally modulated Josephson critical current}
\author{A. E. Koshelev}
\affiliation{Materials Science Division, Argonne National Laboratory, Argonne, Illinois 60439}
\author{L. N. Bulaevskii}
\affiliation{Los Alamos National Laboratory, Los Alamos, New Mexico 87545}
\date{\today }

\begin{abstract}
Intrinsic Josephson-junction stacks realized in mesas fabricated out of
high-temperature superconductors may be used as sources of coherent
electromagnetic radiation in the terahertz range. The major challenge is to
synchronize Josephson oscillations in all junctions in the stack to get
significant radiation out of the crystal edge parallel to the $c$ axis. We
suggest a simple way to solve this problem via artificially prepared lateral
modulation of the Josephson critical current identical in all junctions. In
such a stack phase oscillations excite the in-phase Fiske mode when the
Josephson frequency matches the Fiske-resonance frequency which is set by the
stack lateral size. The powerful almost standing electromagnetic wave is
excited inside the crystal in the resonance. This wave is homogeneous across
the layers meaning that the oscillations are synchronized in all junctions in
the stack. We evaluate behavior of the I-V characteristics and radiated power
near the resonance for arbitrary modulation and find exact solutions for
several special cases corresponding to symmetric and asymmetric modulations of
the critical current.
\end{abstract}
\maketitle

\section{Introduction}

Josephson junctions are natural voltage-to-frequency converters, since a finite
voltage drop across the junction always leads to oscillating current with
frequency proportional to the voltage (\textit{ac} Josephson effect
\cite{JosPL62}). This fundamental property suggested an attractive possibility
to use the ac Josephson effect for developing of voltage-tunable generators of
electromagnetic waves. Radiation from a Josephson junction directly into the
waveguide in the microwave frequency range has been indeed detected a long time
ago \cite{Dm65,LanPRL65}. However, the typical detected radiated power $\sim$1\
pW occurred to be too small for practical applications.

A natural route to enhance this power is to use large arrays of junctions. This
possibility has been extensively explored by several experimental groups, see
reviews \cite{Jain,Darula99}. When all junctions oscillate in phase, the total
emitted power is expected to be proportional to the square of the total number
of junctions in the array. Inevitable variations of junction parameters,
however, may cause variations of the oscillating frequencies leading to
desynchronization and dramatic drop in emission power. Therefore, the major
challenge is to synchronize oscillations in all junctions \cite{Jain}. One way
to solve this problem is to couple the junctions with a resonant cavity. The
efficient synchronization in such systems has been demonstrated experimentally
\cite{BarbaraPRL99} and has been extensively studied in several simulation
papers \cite{StroudPRB,Filatrella}.

Demonstration of intrinsic Josephson effect in high-temperature superconductors
\cite{KleinerPRL92} opened a completely new route to developing electromagnetic
sources. Layered high-temperature superconducting materials, such as
Bi$_{2}$Sr$_{2}$CaCu$_{2}$O$_{8}$ (BSCCO), are composed of superconducting
CuO$_{2}$ layers coupled by Josephson interaction. Intrinsic Josephson effect
has been extensively investigated in the past decade and most ``classical''
\textit{ac} and \textit{dc} Josephson phenomena have been observed, including
Fraunhofer magnetic oscillations of critical current in small-size samples
\cite{LatyshPRL96}, Josephson plasma resonance \cite{TsuiPRL96,Matsuda2},
Shapiro steps in current-voltage characteristics induced by external microwave
irradiation \cite{WangPRL01,LatyshevPRL01}, Fiske resonances
\cite{IriePL98,KrasnovPRB99,KimPRB04}, etc. Therefore a small-size mesa
fabricated out of this material represents a natural one-dimensional array of
Josephson junctions. A large value of superconducting gap, up to 60 meV, allows
to bring the Josephson frequency into the practically important terahertz
range. Due to atomic nature of the intrinsic junctions, it is much easier to
prepare large arrays of virtually identical junctions than in the case of
artificially fabricated arrays. Nevertheless, the same challenge to synchronize
oscillations in all junctions also remains for this system.

Electromagnetic waves propagate inside large-size layered superconductor in the
form of plasma modes. In zero magnetic field the minimum frequency of these
waves corresponding to homogeneous oscillations is given by the Josephson
plasma frequency. The in-plane velocity of the mode strongly depends on the
c-axis wave vector, $q_{z}$. The highest velocity corresponds to the in-phase
mode, $q_{z}=0$, and the lowest velocity corresponds to the antiphase mode,
$q_{z}=\pi/s$.

A stack of the intrinsic Josephson junctions with lateral size smaller than the
decay length of electromagnetic wave behaves as a cavity. It is characterized
by a spectrum of Fiske resonant modes corresponding to excitation of almost
standing electromagnetic waves \cite{KleinerPRB94}. Frequencies of these modes
strongly depend on the wave vector perpendicular to the layer direction, with
the maximum frequency corresponding to the homogeneous in-phase mode in all
junctions and the minimum frequency corresponding to the antiphase mode. At the
stack edge the electromagnetic waves excited inside the intrinsic junctions are
converted into electromagnetic waves propagating into dielectric media outside
the crystal. To use the stack as a coherent source of such radiation, it would
be desirable to excite the in-phase resonance mode. Then the radiation power is
proportional to the square of the total number of junctions positioned at
distances smaller than the wavelength of radiation. However, to synchronize the
intrinsic junctions one needs to have strong enough coupling between them. In
the simplest case of a homogeneous stack at zero magnetic field, the Josephson
oscillations typically interact very weakly inside the crystal.

A very popular way to achieve coherent Josephson oscillations in the whole
stack is to apply magnetic field along the layers. The magnetic field promotes
strong inductive interaction between the neighboring junctions. Large magnetic
field generates a dense Josephson vortex (JV) lattice homogeneously filling all
junctions. In the case of large-size system, interaction between the static JV
arrays in neighboring junctions leads to formation of the triangular lattice
corresponding to the $\pi$ phase shift between the phases in the neighboring
junctions. The Fiske resonances excited by the moving JV lattice have been
observed experimentally \cite{IriePL98,KrasnovPRB99,KimPRB04}. Due to its
triangular ground state, the JV lattice typically excites the antiphase modes.
Only a very weak outside radiation at double Josephson frequency is expected in
this case.\cite{BulKoshGinzb06}

To achieve a powerful emission it would be desirable to prepare aligned
rectangular arrangement of JVs. Such configuration is expected at certain
conditions in small-size mesas due to interaction with edges. At small lattice
velocities transition from triangular to rectangular configuration with
increasing magnetic field has been observed experimentally as a crossover from
$\Phi_{0}/2$- to $\Phi_{0}$-periodic magnetic oscillations of the flux-flow
voltage \cite{KakeyaMagOsc}. The transitions between \emph{static}
configurations have been studied theoretically in Ref.\ \onlinecite{AEKMagOsc}.
The possibility to have the aligned configuration at large velocities is an
open issue. Recent large-scale numerical simulations \cite{TachikiPRB05}
suggest that excitation of the resonance in-phase mode and interaction via the
radiation electromagnetic field may promote alignment of JVs.

A possibility to use a mesa with small lateral size and a very large number of
junctions (about 10$^4$) in zero magnetic field as a source of terahertz
radiation has been proposed in Ref.\ \onlinecite{BulKoshPRL07}. In such a
design oscillations in different junctions are synchronized by the external
electromagnetic radiation field generated by the oscillations themselves. Small
lateral size increases the strength of interaction due to the radiation field
and allows to avoid excessive heating. Such a source is frequency-tunable with
the maximum power conversion efficiency about 30\%. The obvious technological
challenge of this design is a requirement to fabricate a mesa with such a large
number of almost identical junctions.

In this paper we explore a different way to excite resonance mode and
synchronize oscillations in a junction stack. We propose to use a stack with
strongly modulated Josephson critical current (JCC) with modulation identical
in all junctions. Such a modulation dramatically enhances coupling between
the Josephson oscillations and the in-phase resonance modes. For a single junction
such a mechanism of excitation of the Fiske resonances has been considered in
Ref.\ \onlinecite{RussoVaglio78}.

The frequencies of in-phase resonance modes are set by the lateral size of the
mesa, $L$,
\begin{equation}
\omega_{m}=\frac{c}{\sqrt{\varepsilon_{c}}}\frac{m\pi}{L}.\label{ResModes}
\end{equation}
where $\varepsilon_{c}$ is the c-axis dielectric constant. In particular,
assuming $\varepsilon_{c}=12$, the resonance at $\omega_{1}/2\pi=1 $ THz for
the fundamental mode, $m=1$, takes place for a mesa with width 43 $\mu$m.  In
the resonance a powerful almost standing wave is excited by Josephson
oscillations which synchronizes oscillations in the whole stack. Such
synchronization function of the cavity mode does not exist in a single
junction.\cite{RussoVaglio78}  In such a design the frequency tunability in a
single device is sacrificed in favor of larger power and better efficiency. We
consider several specific cases of modulation, both symmetric and asymmetric,
which correspond to excitation of the mode with the wavelength equal to $L$ and
$2L$. For simplicity, we assume that the mesa size along the field direction,
$L_{y}$, is larger than the wavelength of outcoming electromagnetic wave,
$\lambda_{\omega}$ (0.3 mm for 1 THz). The calculation can be straightforwardly
generalized to the opposite case $L_{y}<\lambda_{\omega}$. We calculate the I-V
dependences, radiated electromagnetic power, and power conversion efficiency
for such systems.

%Resub:
Recently, the resonant electromagnetic emission from the mesas fabricated  out
of underdoped BSCCO crystals has been detected experimentally
\cite{LutfiSci07}. The resonance frequencies vary from 0.4 to 0.85 THz, and
they increase roughly inversely proportional to the mesa widths.
%which vary from 100$\mu$m down to 40$\mu$m.
%The mesas lengths $\sim 300\mu$m were somewhat smaller then the
%wavelength for these frequencies $\sim 600\mu$m, and the mesa heights were
%$\sim 1\mu$m.
%Power levels up to 50 nW in the resonance have been detected by the remote
%bolometer, while power levels pumped into the resonance mode, estimated from
%the resonance features in IV characteristics, can reach $\sim 20\mu$W.
%These experiments demonstrate that the heating effects can be manageable even
%in large-size mesas with lateral sizes of several hundred micrometers in the
%voltage range corresponding to the Josephson frequencies around 1 THz.
The origin of the observed resonances is most probably due to the mechanism
described in this paper, even though no JCC modulation has been introduced
deliberately. Suppression of superconductivity near the edges during the
fabrication process occurs to be sufficient for excitation of the resonances.
One can expect that deliberately introduced JCC modulation will significantly
enhance the amplitude of the resonance and radiation power.

\begin{figure*}[ptb]
\begin{center}
\includegraphics[width=6.in]{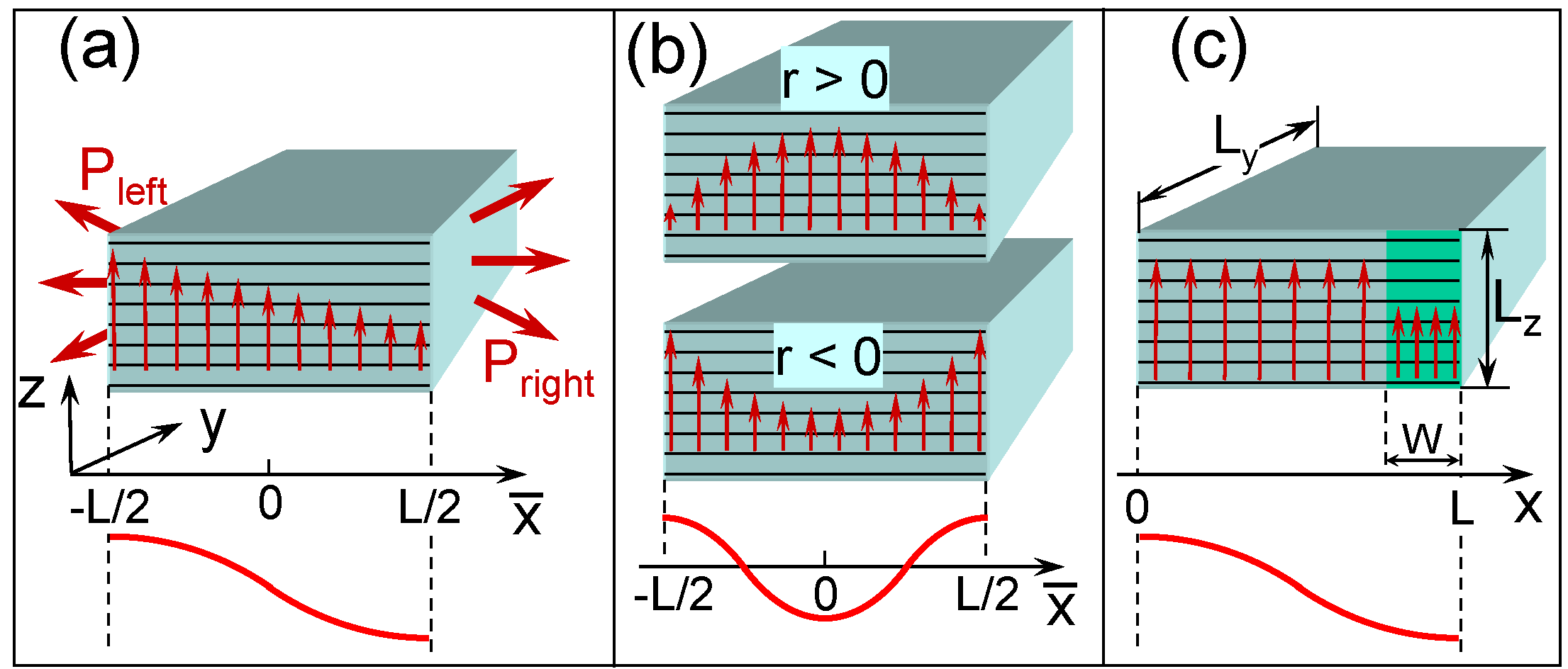}
\end{center}
\caption{(Color online) Mesas with modulation of the Josephson critical current
density considered in the paper: (a) linear modulation (b) parabolic
modulation, and (c) steplike suppression of the critical current near the edge.
The lower plots illustrate the shapes of lowest excited Fiske resonance modes.
} \label{Fig-ModulSchem}
\end{figure*}

The paper is organized as follows. In section \ref{Sec-Gen} we present the
equation and boundary conditions for the oscillating phase when it is identical
in all junctions. Derivation of the boundary condition for this case is
summarized in Appendix \ref{App-Bound}. We also present the radiation power in
terms of the oscillating phase and power conversion efficiency. In section
\ref{Sec-EnBal} we derive the energy balance relations in the dynamic state. In
section \ref{Sec-NearRes}, using these relations, we analyze the behavior near
the resonances and derive approximate results for resonant enhancements of the
current, radiated power, and power conversion efficiency. Appendix
\ref{App-Rad} presents derivation of the radiation losses for the resonance
mode in a thin rectangular mesa. In section \ref{Sec-Cases} we consider several
special cases of modulation for which the problem allows for exact analytical
solutions, see Fig.\ \ref{Fig-ModulSchem}. We perform a detailed analysis of
transport and radiation properties for these cases.

\section{General relations \label{Sec-Gen}}

We consider a stack of intrinsic Josephson junctions (mesa) located at $0<x<L$,
with modulated JCC $j_{J}(x)=g(x)j_{J}$, where $j_{J}$ is the JCC density at
the reference point at which $g(x)=1$. When all junctions oscillate in-phase,
the $c$-axis homogeneous phase obeys the following reduced equation
\begin{equation}
\frac{\partial^{2}\varphi}{\partial\tau^{2}}+\nu_{c}\frac{\partial\varphi
}{\partial\tau}+g(x)\sin\varphi-\frac{\partial^{2}\varphi}{\partial x^{2}
}=0,\label{HomPhaseEq}
\end{equation}
in which the unit of length is the c-axis London penetration depth,
$\lambda_{c}$, and the unit of time is the inverse plasma frequency,
$1/\omega_{p}$. Both $\lambda_{c}$ and $\omega_{p}$ are also defined at the
reference point at which $g(x)=1$. We will use these reduced units throughout
the paper, converting to real units only in some important final results.  The
reduced damping parameter, $\nu_{c}$, is related to the quasiparticle tunneling
conductivity, $\nu_{c}=4\pi\sigma_{c}/\varepsilon_{c}\omega_{p}$. We will
neglect inhomogeneity in the dissipation parameter $\nu_{c} $, because
dissipation plays a minor role in the following consideration.

In the resistive state the phase is given by
\begin{equation}
\varphi=\tilde{\omega}\tau+\theta(\tau,x),\ \theta(\tau,x)=\operatorname{Re}
\left[  \theta_{\omega}(x)\exp(-i\tilde{\omega}\tau)\right] .\label{PhaseResSt}
\end{equation}
where $\tilde{\omega}=\omega/\omega_{p}$ is the reduced Josephson frequency. We
will use the linear approximation for the oscillating phase $\theta(\tau)$
valid for $\theta(\tau)<1$. As $\sin(\tilde{\omega}\tau)=\operatorname{Re}
[i\exp(-i\tilde{\omega}\tau)]$, the amplitude of the oscillating phase,
$\theta_{\omega}$, obeys the following equation
\begin{equation}
\left(  \tilde{\omega}^{2}+i\nu_{c}\tilde{\omega}\right)  \theta_{\omega
}+\frac{\partial^{2}\theta_{\omega}}{\partial x^{2}}=ig(x).\label{OscPhEq}
\end{equation}

The boundary conditions follow from the relation between the oscillating
electric and magnetic fields in outside medium and the Josephson relations
between the oscillating phase and these fields \cite{BulKoshGinzb06}. In
general, the boundary conditions for the $c$-axis homogeneous oscillating phase
describing radiation can be written as
\begin{equation}
\frac{\partial\theta}{\partial x}=\pm\int_{-\infty}^{\tau}d\tau^{\prime}
\beta(\tau-\tau^{\prime})\frac{\partial\theta}{\partial\tau^{\prime}
},~\text{for\ }x=0,\ L\label{BoundCondGen}
\end{equation}
or, in Fourier representation,
\begin{align}
\partial\theta_{\omega}/\partial x  &  =\mp i\zeta\theta_{\omega}
,~\text{for\ }x=0,\ L\label{BoundCond}\\
\text{with }\zeta &  =\tilde{\omega}\beta_{\omega}\text{ and }\beta_{\omega
}=\int_{0}^{\infty}d\tau\beta(\tau)\exp(i\omega\tau).\nonumber
\end{align}
Here the kernel $\beta(\tau-\tau^{\prime})$ depends on electromagnetic
properties of the outside media. These boundary conditions assume that there is
only outcoming waves at both boundaries. This means that we neglect reflected
waves, coming back to the stack and mixture of radiation coming from the
opposite sides. Such a mixture can be suppressed, if the mesa is bounded by
large metallic contacts on both sides acting like screens. Derivation of the
boundary condition in such a situation is presented in Appendix \ref{App-Bound}
and gives the following result for $\zeta(\tilde{\omega})$
%For example, in the simple case of a stack taller than the wave length of
%outside radiation this kernel is simply determined by the speed of light in the
%media $c_{0}$ (in units of $\lambda_{c}\omega_{p} =c/\sqrt{\varepsilon_{c}}$),
%namely, $\beta(\tau-\tau^{\prime})=\delta (\tau-\tau^{\prime})/c_{0}$. In the
%more realistic case of a stack with height $L_{z}=sN$ smaller than the
%wavelength, $k_{\omega}L_{z}\ll1$, the parameter $\beta_{\omega}$ is roughly
%reduced by the factor $k_{\omega}L_{z}$. Here $N$ is the number of junctions in
%the stack, $s$ is the interlayer periodicity, and $k_{\omega}=\omega/c$ is the
%wave vector of outside radiation. In this case for the boundary with free
%space, $\zeta$ approximately is given by \cite{BulKoshGinzb06}
\begin{equation}
\zeta=\tilde{\omega}\beta_{\omega}=\frac{L_{z}\tilde{\omega}}{2\varepsilon
_{c}\lambda_{c}}\left[  |\tilde{\omega}|-\frac{2i\tilde{\omega}}{\pi}\ln
\frac{5.03\sqrt{\varepsilon_{c}}\lambda_{c} }{|\tilde{\omega}|L_{z}}\right]
.\label{betaN}
\end{equation}
The radiation losses are determined by the real part of $\zeta$. Its imaginary
part only slightly displaces the resonance frequencies. In the following, we
will neglect the imaginary part in the analytical estimates.
%A similar problem of radiation out of the rectangular cavity has been
%considered in Ref.\ \onlinecite{LeoneIEEE03}.
Using typical values $N=1000$ and $\lambda_{c}=185\mu$m\cite{ftnt-underdp}, we
estimate $sN/2\varepsilon_{c}\lambda_{c}=3.5\cdot10^{-4}$, indicating that
$|\zeta|$ is typically very small.

The oscillating phase determines transport and radiation properties of the
mesa. Without interference, the total radiation loss, $P_{\mathrm{tot}}$, is a
sum of radiation powers coming from the left and right, sides,
$P_{\mathrm{tot}}(\omega
)=P_{\mathrm{left}}(\omega)+P_{\mathrm{right}}(\omega)$. The left-side power,
$P_{\mathrm{left}}$, is given by Poynting vector at the boundary, which is
determined by the oscillating electric and magnetic fields at this side. These
fields, in turn, may be related to the boundary value of the phase,
$\theta_{\omega}(0)$. In the case of the boundary with free space, in real
units, $P_{\mathrm{left}}(\omega)$ can be presented as \cite{BulKoshGinzb06},
\begin{equation}
P_{\mathrm{left}}(\omega)=L_{y}\frac{\Phi_{0}^{2}\omega^{3}N^{2}}{64\pi
^{3}c^{2}} |\theta_{\omega}(0)|^{2}.\label{poylay}
\end{equation}
Correspondingly, the power radiated from the right side, $P_{\mathrm{right}}$,
is obtained from this formula by replacement $\theta_{\omega}(0)\rightarrow
\theta_{\omega} (L)$. For $\omega/2\pi=1$ THz and $N=1000$, we obtain an
estimate for the prefactor,
\[
\frac{\Phi_{0}^{2}\omega^{3}N^{2}}{64\pi^{3}c^{2}}\approx0.6\frac{\mathrm{W}}{\mathrm{cm}}.
\]
This estimate provides the upper limit for possible radiation power in the
case of strong resonance $\theta_{\omega}\sim1$.

We can also obtain a useful general expression for the power conversion
efficiency, $Q=P_{\mathrm{tot}}/(j_{z}E_{z}L)$, the fraction of the total
power fed to the junction which is converted to radiation. The total power fed
into the stack can be represented as
\begin{equation}
j_{z}E_{z}=\frac{\Phi_{0}^{2}\omega}{s^{2}\pi(4\pi\lambda_{c})^{2}}(\nu
_{c}\tilde{\omega}+i_{J}),\label{TotalPowerGen}
\end{equation}
where $\nu_{c}\tilde{\omega}$ is the quasiparticle current $\sigma_{c} E_{z}$
in units of $j_{J}$ and $i_{J}\equiv\langle g(x)\sin\phi\rangle$ is the reduced
JCC density,
\begin{equation}
i_{J}\approx\frac{1}{2L}\int_{0}^{L}dxg(x)\operatorname{Re}\left[
\theta_{\omega}(x)\right]. \label{JosCurrGen}
\end{equation}
Combining Eqs.\ (\ref{poylay}) and (\ref{TotalPowerGen}), we derive
\begin{equation}
Q =\frac{\operatorname{Re}[\zeta]}{2L}\frac{|\theta_{\omega}(0)|^{2}
+|\theta_{\omega}(L)|^{2}}{\nu_{c}\tilde{\omega}+i_{J}}.
\label{PowerFrac}
\end{equation}
In the following sections, we will first consider the behavior near the
resonance frequencies and then we will present the most interesting special
cases of modulation allowing for exact solutions.

\section{Energy balance \label{Sec-EnBal}}

We consider first the energy-balance relations. The reduced energy in units of
$L_{z}L_{y}\Phi_{0}^{2}/(32\pi^{3}s^{2}\lambda_{c})$ accumulated in the phase
oscillations inside the stack is given by
\begin{equation}
\mathcal{E}=\int_{0}^{L}dx\left[  \frac{1}{2}\left(  \frac{\partial\theta
}{\partial\tau}\right)  ^{2}+\frac{1}{2}\left(  \frac{\partial\theta}{\partial
x}\right)  ^{2}\right]. \label{OscEnergy}
\end{equation}
Therefore, the energy-change rate is given by
\begin{equation}
\frac{\partial\mathcal{E}}{\partial
\tau}=\!\int_{0}^{L}\!dx\frac{\partial\theta}{\partial\tau}\left[
\frac{\partial^{2}\theta}{\partial\tau^{2}}-\frac{\partial^{2}\theta}{\partial
x^{2}}\right]  \!+\!\left[  \frac{\partial\theta}{\partial\tau}\frac
{\partial\theta}{\partial x}\right] _{L}\!-\left[  \frac{\partial\theta
}{\partial\tau}\frac{\partial\theta}{\partial x}\right]  _{0}.\label{EnRate1}
\end{equation}
As follows from the Eq.\ (\ref{HomPhaseEq}), the first term can be transformed
to
\begin{align*}
& \int_{0}^{L}dx\frac{\partial\theta}{\partial\tau}\left[  \frac{\partial
^{2}\theta}{\partial\tau^{2}}-\frac{\partial^{2}\theta}{\partial x^{2}
}\right] \\
& =-\nu_{c}\int_{0}^{L}dx\left(  \frac{\partial\theta}{\partial\tau}\right)
^{2}-\int_{0}^{L}dx\frac{\partial\theta}{\partial\tau}g(x)\sin\tilde{\omega}
\tau.
\end{align*}
Here, the first term accounts for the quasiparticle damping, while the second
term gives the driving force from the Josephson oscillations leading to pumping
of energy from a dc source into the electromagnetic oscillations inside the
stack. The last two terms in Eq.\ (\ref{EnRate1}) account for the radiation
losses at the boundaries. For the general boundary conditions
(\ref{BoundCondGen}) these terms can be transformed as
\[
\left[  \frac{\partial\theta}{\partial\tau}\frac{\partial\theta}{\partial
x}\right]  _{L}-\left[  \frac{\partial\theta}{\partial\tau}\frac
{\partial\theta}{\partial x}\right]  _{0}=-\left[  \frac{\partial\theta
}{\partial\tau}\hat{\beta}\frac{\partial\theta}{\partial\tau}\right]
_{L}-\left[  \frac{\partial\theta}{\partial\tau}\hat{\beta}\frac
{\partial\theta}{\partial\tau}\right]  _{0}
\]
where we introduce a notation for the operator $\hat{\beta}\frac{\partial\theta
}{\partial\tau}=\int_{-\infty}^{\tau}d\tau^{\prime}\beta(\tau-\tau^{\prime
})\frac{\partial\theta}{\partial\tau^{\prime}}$. Therefore, the total
energy-change rate can be written as
\begin{align}
\frac{\partial\mathcal{E}}{\partial \tau}= & -\nu_{c}\int_{0}^{L}dx\left(
\frac{\partial\theta }{\partial\tau}\right)
^{2}-\int_{0}^{L}dx\frac{\partial\theta}{\partial\tau
}g(x)\sin(\tilde{\omega} \tau)\nonumber\\
- & \left[  \frac{\partial\theta}{\partial\tau}\hat{\beta}\frac{\partial
\theta}{\partial\tau}\right]  _{L}-\left[  \frac{\partial\theta}{\partial\tau
}\hat{\beta}\frac{\partial\theta}{\partial\tau}\right]  _{0}.\label{EnRate2}
\end{align}
For a steady state, the energy has to remain constant, meaning that the energy
supplied by the Josephson oscillations has to be exactly compensated by the
quasiparticle and radiation losses.

\section{Behavior near resonances for arbitrary modulation \label{Sec-NearRes}}

In this section, we obtain approximate results for the current and radiation in
the vicinity of the resonance frequency $\tilde{\omega}_{m}=m\pi/L$, where the
phase can be approximated as the corresponding cavity mode
\begin{equation}
\theta(x)\approx\psi\cos(m\pi x/L).\label{CavMode}
\end{equation}
We neglect the small influence of the radiation on the shape of the resonance
mode. In this case, the energy in the mode (\ref{OscEnergy}) and energy-change
rate (\ref{EnRate2}) can be approximated as
\begin{align}
\mathcal{E}  &  \approx\frac{L}{2}\left[  \frac{1}{2}\left(  \frac
{\partial\psi}{\partial\tau}\right)  ^{2}+\frac{1}{2}\tilde{\omega}_{m}
^{2}\psi^{2}\right], \label{OscEnerMode}\\
\frac{\partial\mathcal{E}}{\partial \tau}  &  \approx\!-\frac{L}{2}\left[
\nu_{c}\left(  \frac {\partial\psi}{\partial\tau}\right)
^{2}\!+\frac{\partial\psi}{\partial\tau
}g_{m}\sin\tilde{\omega}\tau+\frac{4}{L} \frac{\partial\psi} {\partial\tau}
\hat{\beta}\frac{\partial\psi}{\partial\tau}  \right], \label{EnRateMode}
\end{align}
where
\begin{equation}
g_{m}=\frac{2}{L}\int_{0}^{L}dx\cos(m\pi x/L)g(x)\label{ModeCouplingConst}
\end{equation}
is the coupling parameter. Therefore, equation for the mode amplitude is given
by
\begin{equation}
\frac{\partial^{2}\psi}{\partial\tau^{2}}+\tilde{\omega}_{m}^{2}\psi+\nu
_{c}\frac{\partial\psi}{\partial\tau}+\frac{4}{L}\hat{\beta}\frac{\partial
\psi}{\partial\tau}=-g_{m}\sin\tilde{\omega}\tau.\label{ModeEquat}
\end{equation}
Using complex representation $\psi=\operatorname{Re}[\psi_{\omega}
\exp(-i\tilde{\omega}\tau)]$, we obtain a solution
\begin{equation}
\psi_{\omega}=\frac{ig_{m}}{\tilde{\omega}^{2}-\tilde{\omega}_{m}^{2}+i\left(
\nu_{c}+\nu_{r}\right)  \tilde{\omega}}.\label{ModeSolutions}
\end{equation}
where
\begin{equation}
\nu_r=\frac{4\beta_{\omega}}{L}=\frac{2L_{z}\omega}{\varepsilon_{c}L\omega_{p}}
\label{RadDampPar}
\end{equation}
is the parameter of the radiation damping (the last formula is written in real
units).  One can see that both the quasiparticle dissipation and radiation
contribute to the resonance damping.\cite{RadDampNote}  The cavity quality
factor is given by $Q_{\mathrm{c}}=\omega_{m}/(\nu_{c}+\nu_{r})$. Optimal power
conversion is achieved when the main contribution to damping is coming from the
radiation, $\nu_{c}\ll\nu_{r}$. Comparing the damping channels using Eq.
(\ref{betaN}), we obtain that this is achieved for a sufficiently large number
of junctions in the stack
\begin{equation}
N>N_{\sigma}=\frac{\varepsilon_{c}\nu_{c}L}{2s\tilde{\omega}}=\frac{2\pi
\sigma_{c}L}{\omega s}.\label{condition}
\end{equation}
Taking $s=1.56$nm, we also rewrite this formula in the practically convenient
form as
\[
N_{\sigma}\approx576\ \sigma_{c}[1/(\Omega\cdot\mathrm{cm})]L[\mu
\mathrm{m}]/f[\mathrm{THz}].
\]
For typical values $\sigma_{c}=0.003-0.01$ ($\Omega\cdot$cm)$^{-1}$,
$L\sim40\mu$m, and $f=\omega/2\pi=1$ THz, we estimate $N_{\sigma}= 70-250$. In
the regime of dominating radiation losses, the cavity quality factor is simply
given by $Q_{\mathrm{c}}=\varepsilon_{c} L/(2 L_z)$.

The solution (\ref{ModeSolutions}) allows us to obtain the average JCC
\begin{align}
i_{J}  &  =\left\langle g(x)\sin\left\{  \tilde{\omega}\tau+\operatorname{Re}
\left[  \psi_{\omega} \exp(-i\tilde{\omega}\tau)\right]  \cos(m\pi x/L)
\right\}  \right\rangle \nonumber\\
&  =\frac{1}{4}\frac{g_{m}^{2}\left(  \nu_{c}+\nu_{r}\right)
\tilde{\omega}}{\left[ \tilde{\omega}^{2}-\tilde{\omega}_{m}^{2}\right]
^{2}+\left( \nu_{c}+\nu_{r}\right)  ^{2}\tilde{\omega}^{2}}.
\label{JosCurrentMode}
\end{align}
This gives the maximum current enhancement in the resonance
\begin{equation}
i_{J,\max}=\frac{g_{m}^{2}/4}{\left(  \nu_{c}+\nu_{r}\right)
\tilde{\omega}_{m}}.\label{JosCurrMax}
\end{equation}
A similar result has been derived in Ref.\ \onlinecite{RussoVaglio78} for the
case of a single junction without radiation losses. Comparing this result with
the reduced quasiparticle current, $\nu_{c} \tilde{\omega}$, we see that the
resonance feature in I-V characteristic is pronounced if $g_{m}>2\sqrt{\left(
\nu_{c}+\nu_{r}\right)  \nu_{c} }\tilde{\omega}$. In the case of dominating
radiation losses we can rewrite this condition in a more transparent form,
$g_{m}>2\sqrt{2L_{z}\nu _{c}/(\varepsilon_{c}L})\tilde{\omega}^{3/2}$. For
$\nu_{c}=0.01$, $\tilde {\omega}=10 $, $L_{z}=1.5\mu$m, and $L=40\mu$m
corresponding to $m=1$, we obtain that the resonance feature in the I-V
dependence becomes strong when the coupling parameter exceeds $0.5$. In the
case of strong resonance, the total current
$i(\tilde{\omega})=\nu_{c}\tilde{\omega}+i_{J}$ nonmonotonically depends on the
Josephson frequency $\tilde{\omega}$ (voltage). In this case, only the
increasing part $di/d\tilde{\omega}>0 $ is stable.

The total radiated power from both sides is given by
\begin{align}
P_{\mathrm{tot}}(\omega)  &  =2P_{\mathrm{sc}}\frac{\partial\psi}{\partial
\tau}\hat{\beta}\frac{\partial\psi}{\partial\tau}=2P_{\mathrm{sc}
}\operatorname{Re}[\beta_{\omega}]\tilde{\omega}^{2}|\psi_{\omega}
|^{2}\nonumber\\
&  =\frac{2P_{\mathrm{sc}}\beta_{\omega}\tilde{\omega}^{2}g_{m}^{2}}{\left[
\tilde{\omega}^{2}-\tilde{\omega}_{m}^{2}\right]  ^{2}+\left(  \nu_{c}
+\nu_{r}\right)  ^{2}\tilde{\omega}^{2}}.\label{RadPowerMode}
\end{align}
Here, the scale of $P_{\mathrm{tot}}$ is given by $P_{\mathrm{sc}}=L_{y}
L_{z}\lambda_{c} E_{p}j_{J}/2$, where $E_{p}=\Phi_{0}\omega_{p}/(2\pi cs)$ is
the electric field corresponding to the plasma frequency. For the maximum
radiated power in the resonance we obtain $P_{\mathrm{tot}}(\omega
_{m})=2P_{\mathrm{sc}}\beta_{\omega}g_{m}^{2}/\left(  \nu_{c}+\nu_{r}\right)
^{2}$. In the regime of dominating radiation losses, $\nu_{c}\ll\nu_{r}$, using
Eq.\ (\ref{betaN}) and $j_{J}=c\Phi _{0}/(8\pi^{2}s\lambda_{c}^{2})$, we obtain
a very simple and universal estimate for maximum total radiated power (in real
units)
\begin{equation}
P_{\mathrm{tot}}(\omega_{m})\approx\frac{\pi L_{y}L^{2}g_{m}^{2}j_{J}^{2}}
{2\omega}.\label{RadPowGenReson}
\end{equation}
An important observation is that \emph{for a tall stack in resonance, the
radiated power does not depend on} $N$ due to compensation between the factor
$N^{2}$ in front of the radiated power (\ref{poylay}) and the amplitude of
phase oscillations in the resonance, which, due to the increasing radiation
losses, drops as $ \zeta^{-2}\propto N^{-2}$, see also Ref.\
\onlinecite{BulKoshGinzb06}. This compensation only exists in the regime when
the damping of the resonance is caused by the radiation which is realized under
the condition (\ref{condition}).

The power conversion efficiency is given by
\begin{equation}
Q=\frac{P_{\mathrm{tot}}}{L\left(  i_{J}+\nu_{c}\tilde{\omega}\right)
\tilde{\omega} }.\label{ConvEfficMode}
\end{equation}
In resonance, it can be represented in a quite transparent form as a product of
two factors
\begin{equation}
Q_{r}=\frac{g_{m}^{2}}{g_{m}^{2}+4\left(  \nu_{c}+\nu_{r}\right)
\nu_{c}\tilde{\omega}^{2}}\frac{\nu_{r}}{\nu_{c}+\nu_{r}},\label{ConvEffRes}
\end{equation}
where the first factor represents the relative current increase in the
resonance, $i_{J,\max}/(\nu_c \tilde{\omega}+ i_{J,\max})$, and the second
factor is the relative contribution of the radiation to the resonance damping.
We can see that, remarkably, in resonance the conversion efficiency can
approach $100\%$ provided (i) the resonance feature is pronounced in I-V
dependence, $g_{m}>2\sqrt{\left( \nu_{c} +\nu_{r}\right) \nu_{c}}\tilde{\omega
}$ and (ii) the losses are dominated by the radiation, $\nu_{c} < \nu_{r}$.
Both conditions are quite realistic.

\begin{figure*}[ptb]
\begin{center}
\includegraphics[width=7in]{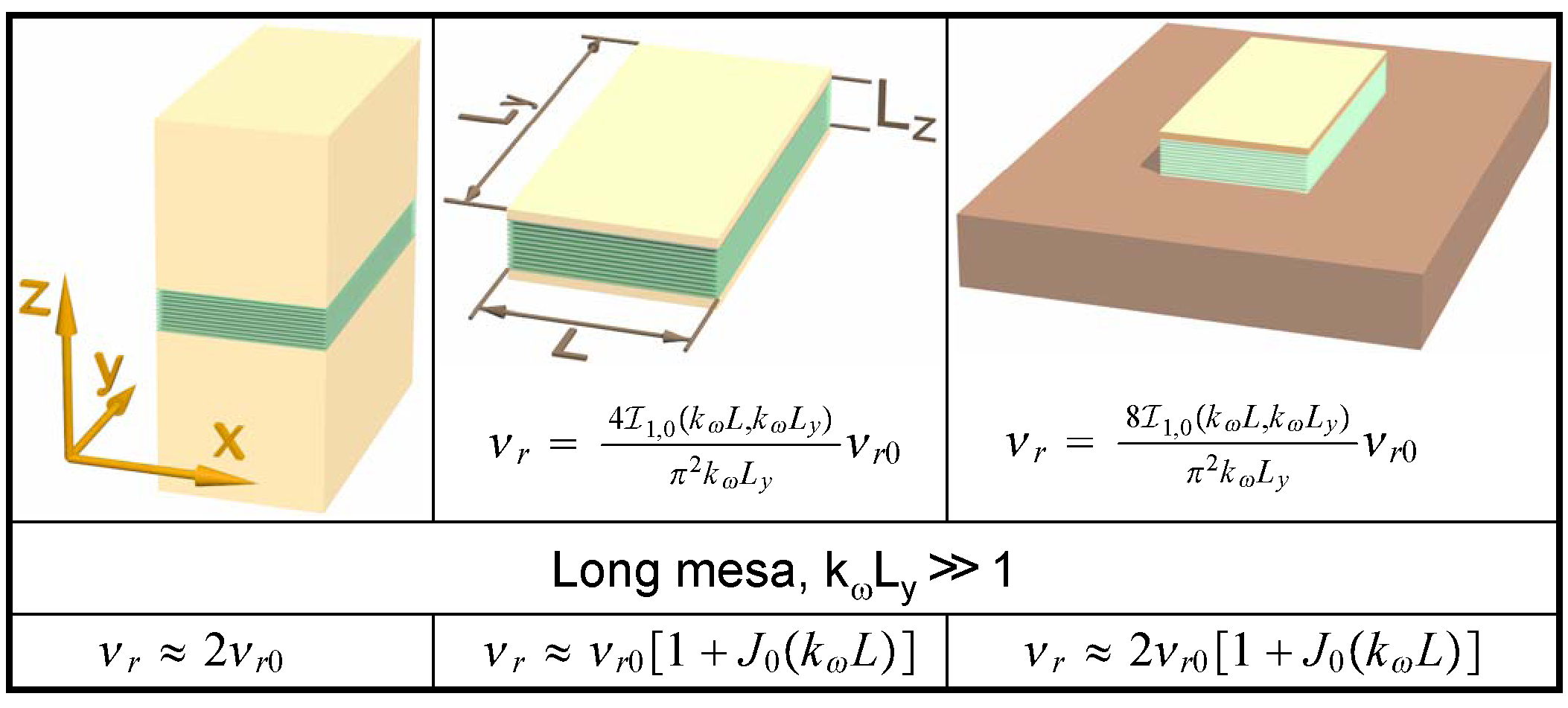}
\end{center}
\caption{(Color online) The parameter of radiation damping for the fundamental
mode for different mesa designs in the case $k_{\omega}L_z\ll 1$: long mesa
with screens mostly considered in the paper (left), rectangular capacitor
(middle), and mesa with ground plate (e.g., mesa fabricated on the top of bulk
crystal) (right). Here $\nu_{r0}=\omega L_{z}/(\varepsilon_{c}\omega_{p}L)$ and
the function $\mathcal{I}_{1,0}(a_{x},a_{y})$ is defined in Appendix
\ref{App-Rad}, Eq.\ (\ref{I10}). In the regime of dominating radiation losses,
the cavity quality factor $Q_{c}$ is directly determined by $\nu_r$ as
$Q_{c}=\omega/(\nu_{r}\omega_{p})$.} \label{Fig-MesaCases}
\end{figure*}
Remember that simple and transparent results for the typical number of
junctions (\ref{condition}), current (\ref{JosCurrMax}), radiation power
(\ref{RadPowGenReson}), and conversion efficiency (\ref{ConvEffRes}) are valid
only in the case $\omega L_y/c>1$ and without mixing of radiation coming from
the opposite sides.  These results can be generalized for other cases.
Radiation losses of the resonance mode in the short rectangular mesa can be
approximately calculated similarly to radiation out of a rectangular
capacitor.\cite{LeoneIEEE03} These calculations are summarized in Appendix
\ref{App-Rad}, and the results for the radiation damping parameter for
different cases are presented in Fig.\ \ref{Fig-MesaCases}. In the case of long
mesas, $k_{\omega}L_{y}\gg 1$, the radiation damping parameters for different
geometries differ only by numerical factors of order unity. In the opposite
limit, $k_{\omega}L_{y}\ll 1$, $\nu_r$ acquires an additional small factor
$\sim k_{\omega}L_{y}$.

\section{Special cases of modulation \label{Sec-Cases}}

In this section we consider several special cases of modulation for which the
problem allows for exact analytical solution and full analysis of transport and
radiation properties.
%Resub:
We will consider three cases: linear modulation, parabolic modulations, and
steplike suppression of the critical current. Practical ways to prepare such
modulations are suggested in the discussion section \ref{Sec-Disc}.

\subsection{Linear modulation}
\begin{figure*}[ptb]
\begin{center}
\includegraphics[width=5in]{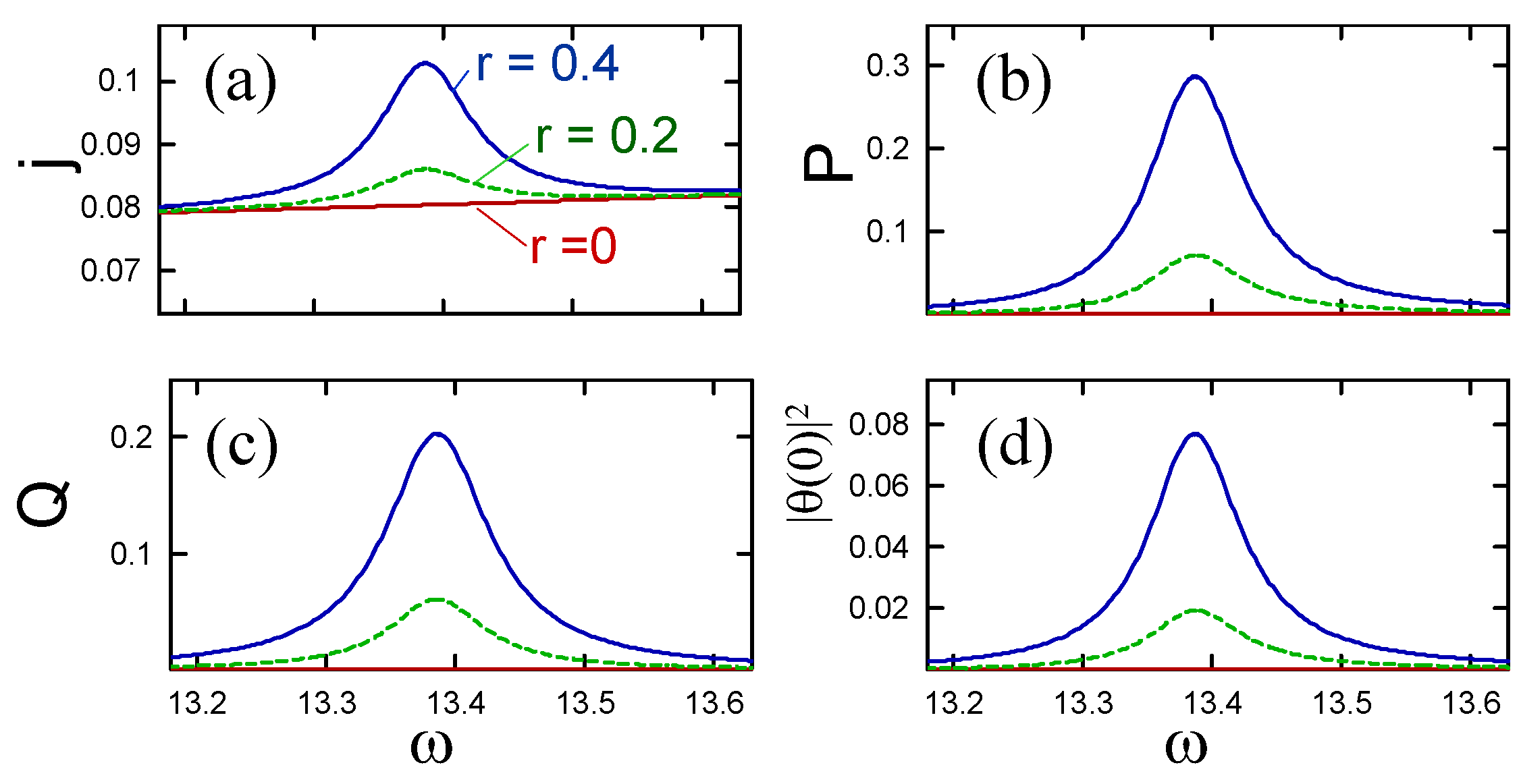}
\end{center}
\caption{(Color online) Representative plots of the Josephson-frequency (or
voltage) dependences of (a) the current density $j$, (b) radiated power from
the left side, $P$, (c) power conversion efficiency $Q$, and (d) amplitude of
oscillating phase at the boundary for stacks with linearly modulated JCC with
parameters $r=0.2$ and $r=0.4$. The unit of current density is the JCC density,
$j_J$, the frequency unit is the plasma frequency $\omega_p$, and the unit of
radiated power is $10^{-3}P_{J}$, see Eq.\ (\ref{PJ}), corresponding to
$P/L_{y}\sim 0.16$W/cm. The following parameters have been used $L=0.23$,
$\nu_{c}=0.006$, and $\mathrm{Re} [\zeta]=0.00035\tilde{\omega}^{2}$. For
comparison, the case of homogeneous JCC ($r=0$) is also shown but $P$, $Q$, and
$|\theta|^{2}$ are indistinguishable from zero at the scales of the plots. In
this case, $P\approx 1.1\cdot 10^{-4}$ and $Q\approx 10^{-4}$ at
$\omega=13.4$.} \label{Fig-LinModulPlots}
\end{figure*}
\begin{figure}[ptb]
\begin{center}
\includegraphics[width=3.2in]{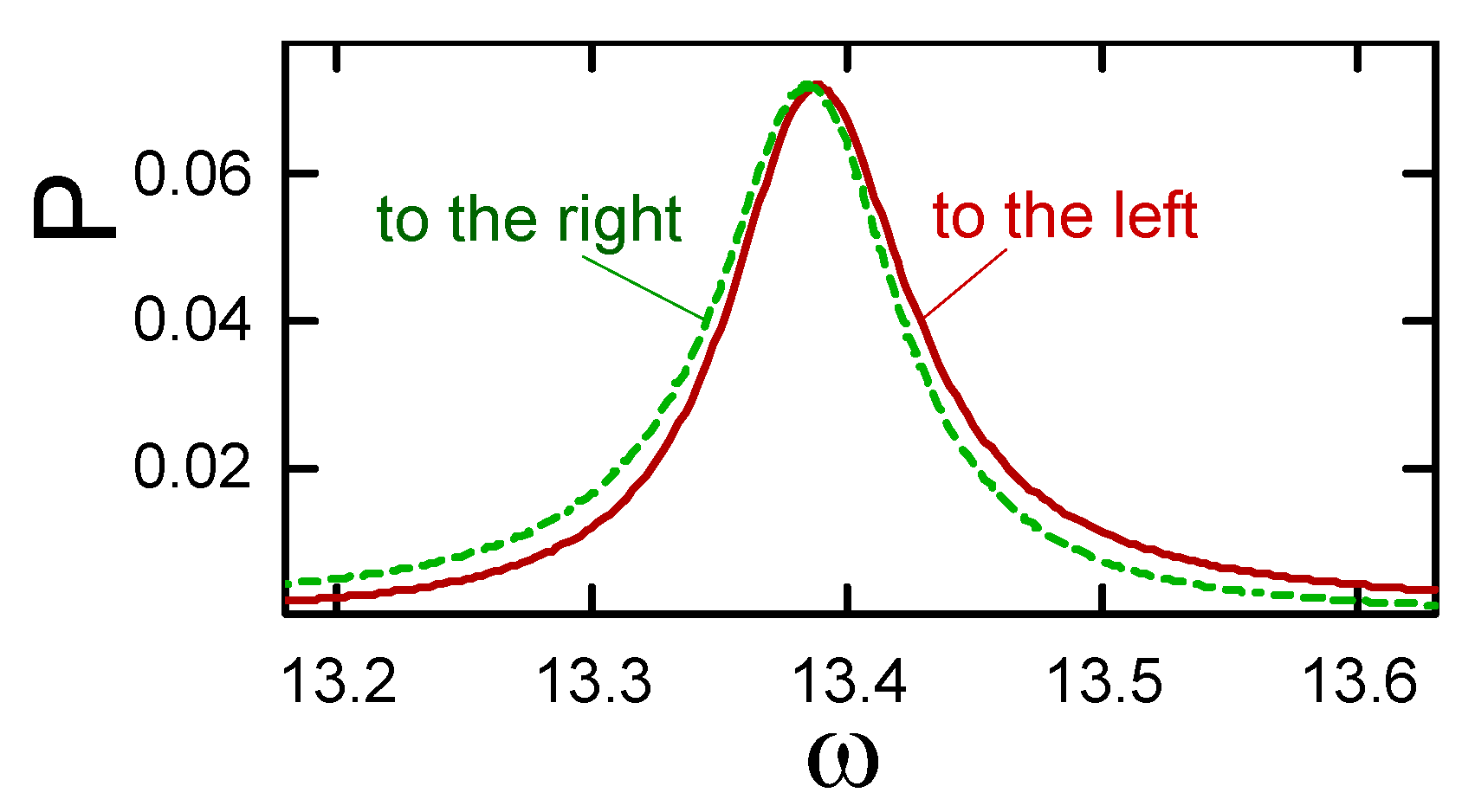}
\end{center}
\caption{(Color online) Comparison between the powers radiated to the two
opposite sides of the mesa for $r=0.2$ and the same parameters as in the
previous figure. One can see that the difference is rather small and amounts to
a slightly different asymmetry of the peaks. } \label{Fig-PowerLinModAsym}
\end{figure}
\begin{figure}[ptb]
\begin{center}
\includegraphics[width=3in]{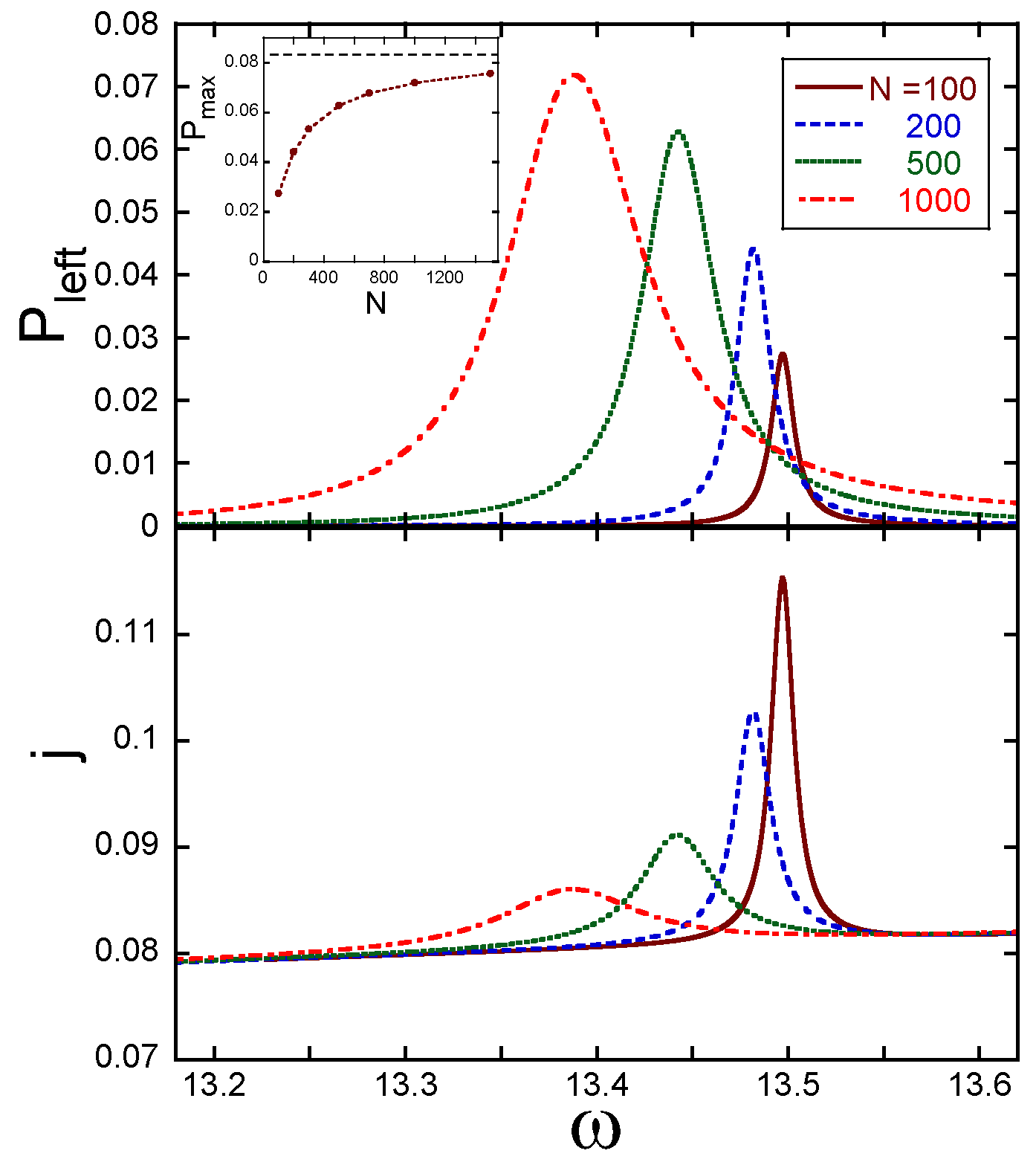}
\end{center}
\caption{(Color online) Evolution of the radiation power  and resonance feature
in the I-V dependence with increasing number of junctions in the stack, $N$,
for $r=0.2$ and the same parameters as in Fig.\ \ref{Fig-LinModulPlots}. Note
that while the radiation power increases with $N$, the resonant feature in
current becomes less pronounced due to the increasing radiation damping
$\propto N^2$. The inset in the upper plot shows the $N$ dependence of the
maximum radiation power, and the dashed line in this plot indicates the
large-$N$ limit obtained from Eq.\ (\ref{RadPowGenReson}).}
\label{Fig-LinModNdep}
\end{figure}

In this section, we consider a mesa, with linearly modulated JCC,
$g(x)=1-2r\bar{x}/L$, see Fig.\ \ref{Fig-ModulSchem}a. Here, we introduce the
new coordinate, $\bar{x}=x-L/2$, which is symmetrical with respect to the mesa,
$-L/2<\bar{x}<L/2$. This means that the JCC at the left side is larger by the
factor $(1+r)/(1-r)$ than the JCC at the right side.
%Resub:
%A mesa fabricated in the region of crystal with inhomogeneous doping would have
%such a modulation. Inhomogeneous doping may be realized in BSCCO crystals with
%oxygen concentration gradient generated by a short-time annealing.
Such a modulation couples homogeneous Josephson oscillations with the odd
Fiske modes (\ref{ResModes}), $m=2l+1$, including the fundamental mode, $m=1$,
and the coupling parameter (\ref{ModeCouplingConst}) to this mode is connected
with the modulation parameter as
\begin{equation}
g_{1}=8r/\pi^{2}.\label{LinModCouplCnst}
\end{equation}

Solution of Eq.\ (\ref{OscPhEq}) with the boundary conditions (\ref{BoundCond})
in the case of linear modulation can be found exactly. Splitting the solution
into the symmetric and antisymmetric parts, $\theta_{\omega}(\bar{x}
)=\theta_{\omega}^{(a)}(\bar{x})+\theta_{\omega}^{(s)}(\bar{x})$, we derive
\begin{align}
\theta_{\omega}^{(a)}(\bar{x})  &  =-\frac{2ir\bar{x}/L}{\tilde{\omega}^{2}
+i\nu_{c}\tilde{\omega}}\nonumber\\
&  +\frac{\left(  2i/L+\zeta\right)  r\sin(p_{\omega}\bar{x})}{\left(
\tilde{\omega}^{2}+i\nu_{c}\tilde{\omega}\right)  \left[  p_{\omega} \cos
(\chi)-i\zeta\sin(\chi)\right]  },\label{LinAntiPhase}\\
\theta_{\omega}^{(s)}(\bar{x})  &  =\frac{i}{\tilde{\omega}^{2}+i\nu_{c}
\tilde{\omega}}\nonumber\\
&  +\frac{\zeta\cos(p_{\omega}\bar{x})}{\left(  \tilde{\omega}^{2}+i\nu
_{c}\tilde{\omega}\right)  \left(  p_{\omega}\sin(\chi)+i\zeta\cos
(\chi)\right)  }\label{LinSymPhase}\\
\text{with }p_{\omega}^{2}  &  \equiv\tilde{\omega}^{2}+i\nu_{c}\tilde{\omega}
\text{ and } \chi\equiv p_{\omega}L/2.\nonumber
\end{align}
Only the antisymmetric phase is coupled to the resonance mode. In particular,
for the boundary phases, we have
\begin{align}
\theta_{\omega}^{(a)}\left(  \pm\frac{L}{2}\right)   &  =\pm\frac{ ir\left[
-\cos(\chi)+\sin(\chi)/\chi\right]  }{p_{\omega} \left[  p_{\omega}\cos
(\chi)\!-\!i\zeta\sin(\chi)\right]  },\label{LinBoundAnti}\\
\theta_{\omega}^{(s)}\left(  \pm\frac{L}{2}\right)   &  =\frac{i \sin(\chi
)}{p_{\omega} \left[  p_{\omega}\sin(\chi)\!+\!i\zeta\cos(\chi)\right]
}.\label{LinBoundSym}
\end{align}
The radiated power is determined by the boundary phases by Eq. (\ref{poylay}),
where we have to replace $\theta_{\omega}(0)$ with $\theta_{\omega}
^{(a)}\left(  \pm L/2\right)  +\theta_{\omega}^{(s)}\left(  \pm L/2\right)  $.

Near the resonance $\tilde{\omega}L=\pi$, using $p_{\omega}\approx
\tilde{\omega}+i\nu_{c}/2$ and $\cos(\chi)\approx(\pi/2-\tilde{\omega}
L/2)-i\nu_{c}L/4$, we obtain
\[
\theta_{\omega}^{(a)}(L/2)\approx-\frac{8ir/\pi^{2}}{\left(  \tilde{\omega}
+i\nu_{c}\right)  \left[ 2(\tilde{\omega}-\pi/L)+i(\nu_{c}+\nu_{r})\right]  },
\]
where $\nu_{r}$ is defined in Eq.\ (\ref{RadDampPar}). Using the coupling
parameter (\ref{LinModCouplCnst}), we can see that this result is consistent
with the general formula (\ref{ModeSolutions}) near the resonance. The maximum
antisymmetric phase in the resonance can be estimated as
$\theta_{\omega}^{(a)}(L/2)\approx-2r/(\pi\beta_{\omega}\tilde{\omega} ^{2})$.
It exceeds the nonresonant symmetric part approximately by the factor
$r/\beta_{\omega}\approx r\varepsilon_{c}L/L_{z}$. In resonance, using Eqs.\
(\ref{RadPowGenReson}) and (\ref{LinModCouplCnst}), we obtain for the radiation
power from one side in real units
\[
P_{r}\approx L_{y}\frac{16r^{2}L^{2}j_{J}^{2}}{\pi^{3}\omega}.
\]
It exceeds nonresonant emission by the factor $(r\varepsilon_{c}L/L_{z})^{2}$.
In particular, for $\omega/2\pi=1$ THz, $j_{J}=500$ A/cm$^{2}$, and $r=0.4$, we
estimate $P_{r}/L_{y} \approx$ 0.05 W/cm.

In the average JCC (\ref{JosCurrGen}), the contributions from symmetric and
antisymmetric parts split, $i_{J}=i_{J,a}+i_{J,s}$. Direct calculation gives
\begin{align}
i_{J,a}  &  =\frac{r^{2}}{2}\left\{  \frac{\nu_{c}/3}{\left(  \tilde{\omega
}^{2}+\nu_{c}^{2}\right)  \tilde{\omega}}\right. \nonumber\\
& \left. +\operatorname{Re}\left[  \frac{\left[  \cos(\chi)-\sin(\chi
)/\chi\right]  \left(  2i/L+\zeta\right) /\chi}{\left(  \tilde{\omega}
^{2}+i\nu_{c}\tilde{\omega}\right)  \left[  p_{\omega}\cos(\chi)-i\zeta
\sin(\chi)\right]  }\right]  \right\}, \label{LinJosCurrAnti}\\
i_{J,s}  &  =\frac{1}{2}\operatorname{Re}\left[  \frac{i}{\tilde{\omega}
^{2}+i\nu_{c}\tilde{\omega}}\left(  1\!-\!\frac{2i\zeta\sin(\chi)/L}
{p_{\omega}\sin(\chi)\!+\!i\zeta\cos(\chi)}\right)  \right].
\label{LinJosCurrSym}
\end{align}
From the general formula for $i_{J,a}$ near the resonance we obtain a much
simpler result
\[
i_{J,a}\approx\frac{16r^{2}}{\pi^{4}}\frac{\left( \nu_{c}+\nu_{r}\right)
/\tilde{\omega}}{ 4(\tilde{\omega}-\pi/L)^{2}+\left(  \nu _{c}+\nu_{r}\right)
^{2} },
\]
and the maximum current enhancement in the resonance is given by
\[
i_{J,a}\approx\frac{16r^{2}/\pi^{4}}{\tilde{\omega}\left(  \nu_{c}
+\nu_{r}\right) }.
\]
These results are also consistent with the corresponding general formulas
(\ref{JosCurrentMode}) and (\ref{JosCurrMax}) if we use the coupling parameter
(\ref{LinModCouplCnst}). For comparison, the symmetric part of the JCC at the
resonance frequency can be estimated as
\[
i_{J,s}\approx\frac{\nu_{c}}{2\tilde{\omega}^{3}}+\frac{\beta_{\omega} }
{\pi\tilde{\omega}^{2}}.
\]
As expected, the resonant enhancement of the current exceeds the nonresonant
radiation correction by the same factor $(r/\beta_{\omega})^{2}=(r\varepsilon
_{c}L/L_{z})^{2}$ as for the radiation power.

For illustration, we present the behavior near the resonance for mesas with two
modulation parameters, $r=0.2$ and $r=0.4$. As a unit of the radiation power,
we selected the quantity
\begin{equation}
P_{J}=L_{y}\lambda_{c}^{2}j_{J}^{2}/\omega_{p}, \label{PJ}
\end{equation}
which is independent on the sizes $L$ and $L_{z}$. This choice of unit is
suggested by the result (\ref{RadPowGenReson}). For $\lambda_c=185$ $\mu$m
$P_{J}/L_y\approx 163$ W/cm. Figure \ref{Fig-LinModulPlots} shows the
Josephson-frequency dependences of (i) the current density $j$ (in units of the
JCC density in the center), (ii) radiated power $P$ (in units of
$10^{-3}P_{J}$), (iii) the power conversion efficiency, $Q$, and (iv) the
amplitude of oscillating phase at the boundary. For comparison, the case of
homogeneous mesa ($r=0$) is also shown. In calculation we used the following
parameters: $L=0.23$, $\nu_{c}=0.006$, and
$\mathrm{Re}[\zeta]=0.00035\tilde{\omega}^{2}$ (corresponding to
$N\approx1000$, $\lambda_{c}\approx 185\ \mu$m, and $\sigma_c\approx0.003$
[$\Omega$ cm]$^{-1}$). We can see that the modulation leads to the appearance
of a strong resonance feature in the I-V dependence. Note that only I-V regions
with positive differential resistivity are stable. Current enhancement in the
resonance is mainly caused by the generation of the powerful electromagnetic
wave and it is accompanied by a huge enhancement of outside radiation.  The
maximum radiation power for used parameters for the case $r=0.4$ corresponds to
$\sim 0.05$W/cm and it exceeds the nonresonant radiation from the homogeneous
mesa by more than 3 orders of magnitude. It is important to note that the power
conversion efficiency is also strongly enhanced in the resonance, reaching
~20\% for $r=0.4$. The plot of $|\theta|^{2}$ shows that for selected
parameters it remains smaller than one in the resonance and, therefore, the
linear approximation used in calculations is not violated.

In spite of the asymmetry of the JCC, the powers radiated to the opposite sides
of the mesa in resonance are approximately the same, because the radiation is
mostly promoted by the resonance mode which has identical amplitudes of the
oscillating electric field at the opposite sides. This is illustrated in Fig.\
\ref{Fig-PowerLinModAsym}, where these powers are plotted for $r=0.2$. One can
see that the peaks have slightly different asymmetries originating from the
symmetric phase.

Figure \ref{Fig-LinModNdep} illustrates the evolution of the radiation power
and resonance feature in the I-V dependence with increasing number of junctions
in the stack, $N$, for $r=0.2$ and the same parameters as in Fig.\
\ref{Fig-LinModulPlots}. The number of junctions above which the radiation
losses dominate (\ref{condition}) can be estimated for used parameters as
$N_{\sigma}\approx 75$. We can see that the current and radiation have opposite
tendencies:  while the radiation power increases with $N$, the resonant feature
in current becomes less pronounced due to the increasing radiation damping.

\subsection{Symmetric parabolic modulation}

\begin{figure*}[ptb]
\begin{center}
\includegraphics[width=5in]{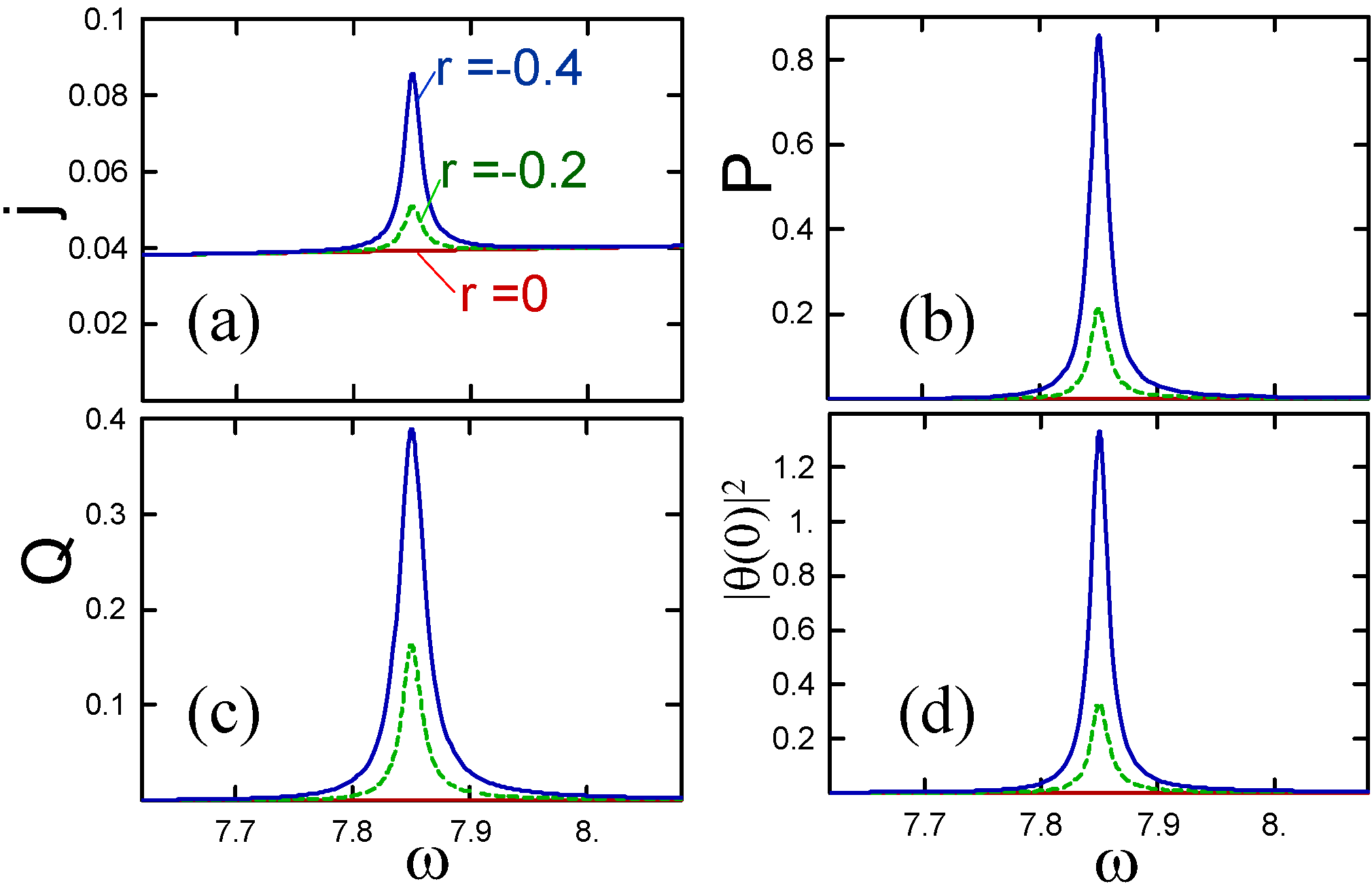}
\end{center}
\caption{(Color online) The Josephson-frequency (or voltage) dependences of (a)
the current density $j$, (b) radiated power to one side $P$, (c) power
conversion efficiency $Q$, and (d) amplitude of the oscillating phase at the
boundary for stacks with parabolic profiles of the JCC and negative modulation
parameters, $r=-0.2$ and $r=-0.4$, corresponding to the case of current
enhancement at the edges.  All units are the same as in Fig.\
\ref{Fig-LinModulPlots}. The following parameters have been used $L=0.8$,
$\nu_{c}=0.005$, and $\mathrm{Re}[\zeta]=0.003\tilde{\omega}^{2}$,
corresponding to $N\approx1000$ and $\lambda_c\approx200\mu$m in the middle.
For comparison, the case of a homogeneous mesa ($r=0$) is also shown. In this
case, $P\approx 5\cdot 10^{-5}$ and $Q\approx 5\cdot 10^{-5}$ at $\omega
=7.85$}. \label{Fig-SymPlots}
\end{figure*}

In this section we consider a symmetric modulation.
%Resub:
%In the case of BSCCO such a modulation can be naturally prepared by short-time
%annealing of the mesa leading to inhomogeneous profile of oxygen concentration.
For simplicity, we assume a simple parabolic profile of the JCC density,
$g(\bar{x})=1-r(2\bar{x}/L)^{2}$, see Fig.\ \ref{Fig-ModulSchem}b, where,
again, $\bar{x}=x-L/2$ and $1-r$ is the ratio of JCCs at the edge and in the
center. The cases $r>0$ and $r<0$ correspond to current suppression and
enhancement at the edges respectively. Such a modulation will lead to
excitation of only even frequency modes (\ref{ResModes}), $m=2l$. In the
following, we will focus on the lowest even mode with $m=2$. To have this
resonance at $\omega_{2}/2\pi=1$ THz assuming $\varepsilon_{c}\approx12$, the
mesa size has to be rather large, $L=86.5$ $\mu$m. The coupling parameter
(\ref{ModeCouplingConst}) to this mode in our case is given by
\begin{equation}
g_{2}=-4r/\pi^{2}.\label{ParabCouplConst}
\end{equation}

The oscillating phase in the case of parabolic modulation also can be found
exactly. From symmetry, the solution of Eq.\ (\ref{OscPhEq}) with boundary
conditions (\ref{BoundCond}) must be an even function of $\bar{x}$, and it has
the following form
\begin{equation}
\theta_{\omega} =\frac{2ir(2/L)^{2}}{\left(  \tilde{\omega}^{2}+i\nu_{c}
\tilde{\omega}\right)  ^{2}}+\frac{i(1-r(2\bar{x}/L)^{2})}{\tilde{\omega}
^{2}+i\nu_{c}\tilde{\omega}}+C\cos(p_{\omega}\bar{x}),\label{OscPhaseSol}
\end{equation}
where the first two terms give a particular solution of the inhomogeneous
equation and the last term is the solution of the homogeneous equations. From
the boundary conditions (\ref{BoundCond}), we obtain
\begin{equation}
C=\frac{-4ir/L+\zeta\left(  2r/\chi^{2} +1-r\right)  }{\left[  p_{\omega}
\sin(\chi)+i\zeta\cos(\chi)\right]  \left(  \tilde{\omega}^{2}+i\nu_{c}
\tilde{\omega}\right)  },\label{ConstSol}
\end{equation}
where $p_{\omega}$ and $\chi$ are defined after Eq.\ (\ref{LinSymPhase}). The
oscillating phase given by Eqs.\ (\ref{OscPhaseSol}) and (\ref{ConstSol})
determines all other observable properties.

The powers radiated from both sides (\ref{poylay}) are determined by the
boundary phase $\theta_{\omega,0 }=\theta_{\omega}(\bar{x}=\pm L/2)$, which we
obtain from Eqs.\ (\ref{OscPhaseSol}) and (\ref{ConstSol}),
\begin{align*}
\theta_{\omega,0} & =\frac{2ir(2/L)^{2}}{\left(  \tilde{\omega}^{2}+i\nu
_{c}\tilde{\omega}\right)  ^{2}}+\frac{i(1-r)}{\tilde{\omega}^{2}+i\nu
_{c}\tilde{\omega}}\\ & +\frac {\left[  -4ir/L+\zeta\left(
2r/\chi^{2}+1-r\right)  \right]  \cos(\chi)}{\left(
p_{\omega}\sin(\chi)+i\zeta\cos(\chi)\right)  \left(  \tilde{
\omega}^{2}+i\nu_{c}\tilde{\omega}\right)  }.
\end{align*}
In the resonance, $\tilde{\omega}L=2\pi$, using $\theta_{\omega}(L/2)\approx
-2r/(\pi\zeta\tilde{\omega})$, we obtain for the  power radiated from one side
in real units
\begin{equation}
P(\omega_{2})\approx\frac{4L_{y}L^{2}r^{2}j_{J}^{2}}{\pi
^{3}\omega}.\label{PoyntRes}
\end{equation}
This result is consistent with the general formula (\ref{RadPowGenReson}) with
the coupling parameter (\ref{ParabCouplConst}). For $\omega/2\pi=1$ THz, $j_{c}
=500$A/cm$^{2}$ and $r=\pm0.5$, we obtain an estimate for the radiated power in
the resonance, $P/L_{y}\approx0.085$W/cm.  Note that $j_{J}$ in Eq.\
(\ref{PoyntRes}) is the JCC density in the center while the the JCC density at
the edge, $j_{J,e}$, is given by $j_{J,e} =(1-r)j_{J}$. In the case of $r<0$
vanishing of superconductivity in the middle, which corresponds to the limits
$j_{J}\rightarrow0$ and $r\rightarrow -\infty$, does not lead to vanishing of
radiation power because $r^{2} j_{J}^{2}\rightarrow j_{J,e}^{2}$ and the
radiation is determined by the JCC density at the edge.

To find the voltage-current characteristic, we calculate the average reduced
JCC (\ref{JosCurrGen}). Using oscillating phase (\ref{OscPhaseSol}), we obtain
\begin{align}
i_{J}& = \frac{\nu _{c}}{2\tilde{ \omega}\left( \tilde{\omega}^{2}+\nu
_{c}^{2}\right) } \left[ 1-\frac{2r}{3}+ \frac{r^{2}}{5}+\frac{16r/L^{2}}{
\tilde{\omega}^{2}+\nu
_{c}^{2}}\left( 1-\frac{r}{3}\right) \right]    \nonumber \\
& +\operatorname{Re}\left\{ \frac{C}{2}\left( \frac{\sin \chi }{\chi
}\!-\!2r\left[ \left( 1\!-\!\frac{2}{\chi ^{2}}\right) \frac{\sin \chi }{2\chi
}+\frac{\cos \chi }{\chi ^{2}}\right] \right) \right\}.   \label{JosCurr}
\end{align}
At the resonance frequency, $\tilde{\omega}L=2\pi$, we estimate $i_{J}
\approx\varepsilon_{c}r^{2}L^{3}/(2\pi^{6}L_{z})$.

Figure \ref{Fig-SymPlots} shows the representative Josephson-frequency
dependences of the current density $j$, radiated power $P$ to one side, the
power conversion efficiency, $Q$, and the amplitude of oscillating phase at the
boundary for negative modulation parameters, $r=-0.2$ and $r=-0.4$,
corresponding to the case of stronger superconductivity at the edges.  In
calculation, we used the following parameters: $L=0.8$, $\nu_{c}=0.005$, and
$\mathrm{Re}[\zeta]=0.003\tilde {\omega}^{2}$ (corresponding to $N\approx1000$
and $\lambda_c\approx 200\mu$m in the center). Overall, the behavior is very
similar to the case of linear modulation shown in Fig.\ \ref{Fig-LinModulPlots}
with minor quantitative differences. We also see that the modulation leads to
the appearance of a strong resonance feature in the I-V dependence accompanied
by a huge enhancement of the outside radiation and power conversion efficiency.

%Resub:
%In the case of high-temperature superconductors there is a straightforward way
%to prepare mesas with symmetric JCC density modulation. The Josephson
%interlayer coupling is very sensitive to the hole doping. This doping can be
%changed in a wide limits by changing oxygen concentration via different kinds
%of annealing at elevated temperatures. Therefore mesas with symmetric modulated
%JCC can be prepared just by short-time annealing. As the oxygen diffusivity in
%BSCCO is strongly temperature dependent \cite{YangAPL99}, the desired
%oxygen-concentration profile can be prepared carefully selecting annealing
%temperature and time.

\subsection{Steplike suppression of critical current near the edge}

\begin{figure}[ptb]
\begin{center}
\includegraphics[width=3.in]{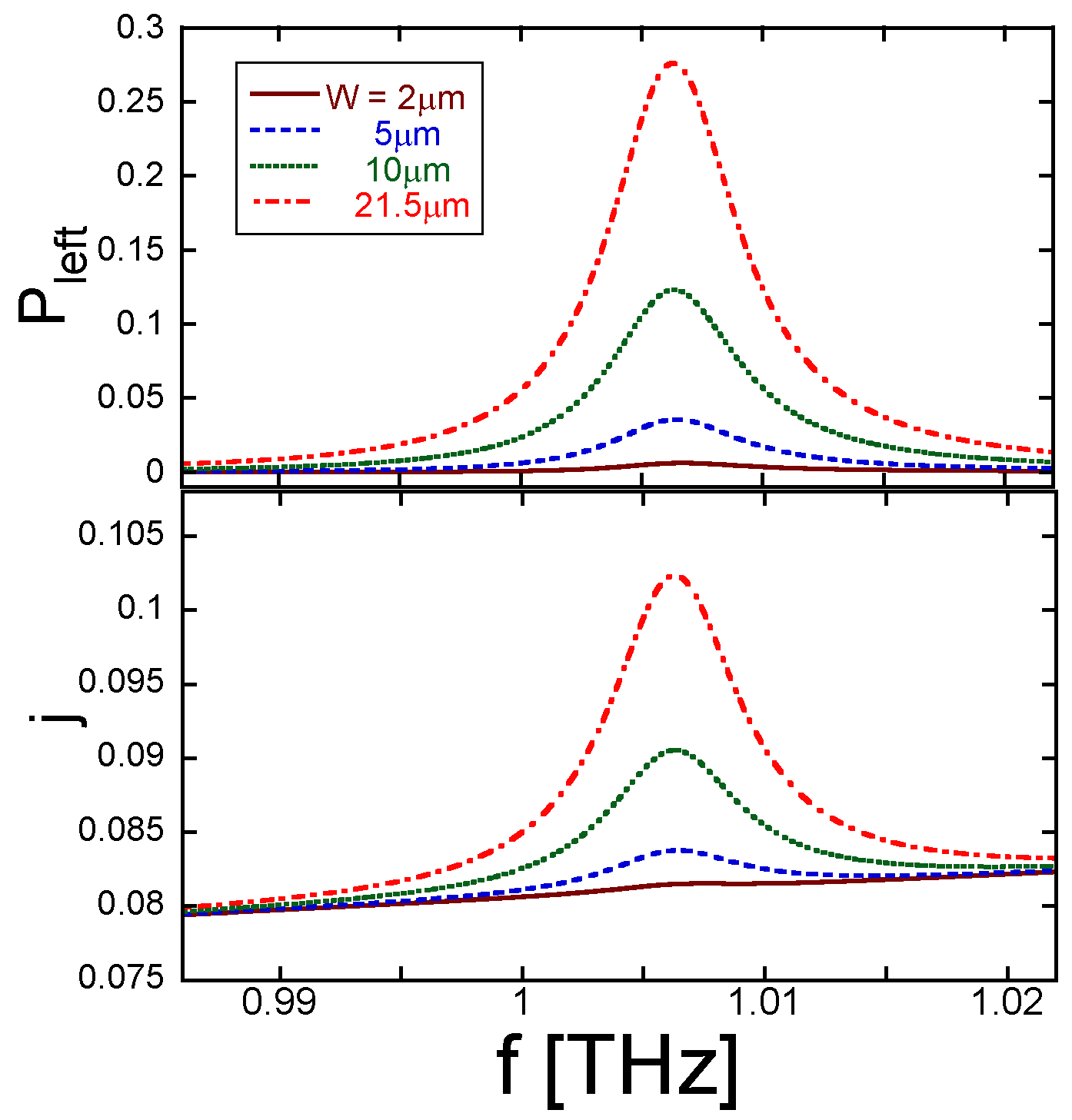}
\end{center}
\caption{(Color online) The radiated power and I-V dependence near the
resonance for different widths of the suppressed region near the mesa right
side. The JCC density in the suppressed region is assumed to be half of its
value in the rest part, $r=0.5$. All units and parameters are the same as in
Fig.\ \ref{Fig-LinModulPlots}}. \label{Fig-StepModWdep}
\end{figure}
In this section, we consider the case when there is a region with suppressed
JCC on one side, see Fig.\ \ref{Fig-ModulSchem}c
\begin{equation}
g(x)=\genfrac{\{}{.}{0pt}{}{1,~\text{for }0<x<L-W}{1-r,~\text{for }L-W<x<L}.
\label{StepModulat}
\end{equation}
%Resub:
%Such a modulation can be generated by the radiation of one side of the mesa
%with protons or electrons.
The coupling parameter (\ref{ModeCouplingConst}) to
the fundamental mode in this case is given by
\begin{equation}
g_{1}=\frac{2r}{\pi}\sin(\pi W/L).\label{StepCouplConst}
\end{equation}

The solution of equation for the oscillating phase (\ref{OscPhEq}) can be built
in the piecewise form,
\begin{equation}
\theta_{\omega}\!=\!\left\{\!
\begin{array}
[c]{l}%
i/(\tilde{\omega}^{2}\!+\!i\nu_{c}\tilde{\omega})+A_{+}\exp(ip_{\omega}x)\\
\ +A_{-}\exp(-ip_{\omega}x),\text{ for }0<x<L-W\\
\\
i(1-r)/(\tilde{\omega}^{2}\!+\!i\nu_{c}\tilde{\omega})+\left(
A_{+}\!-\!C_{+}\right)
\exp(ip_{\omega}x)\\
\ +\left(  A_{-}\!-\!C_{-}\right)  \exp(-ip_{\omega}x),\text{ for
}L\!-\!W\!<\!x\!<\!L
\end{array}
\right.
\end{equation}
%\begin{widetext}
%\begin{equation}
%\theta_{\omega}(x)=
%\genfrac{\{}{.}{0pt}{}{i/(\tilde{\omega}^{2}+i\nu_{c}\tilde{\omega})
%+A_{+}\exp(ip_{\omega}x)+A_{-}\exp(-ip_{\omega}x)\text{, for }0<x<L-W}%
%{i(1-r)/(\tilde{\omega}^{2}+i\nu_{c}\tilde{\omega})%
%+\left(  A_{+}-C_{+}\right) \exp(ip_{\omega}x)+\left(  A_{-}-C_{-}\right)  \exp
%(-ip_{\omega}x)~\text{, for }L-W<x<L} \label{StepPieceSol}
%\end{equation}
with $p_{\omega}^{2}\equiv\tilde{\omega}^{2}+i\nu_{c}\tilde{\omega}$. Matching
$\theta_{\omega}$ and $d\theta_{\omega}/dx$ at $x=L-W$, we obtain
\begin{equation}
C_{\pm}=\frac{-ir/2}{\tilde{\omega}^{2}+i\nu_{c}\tilde{\omega}}\exp\left[  \mp
ip_{\omega}\left(  L-W\right)  \right]
\end{equation}
Using this result, from the boundary conditions (\ref{BoundCond}) we find the
coefficients $A_{\pm}$,
\begin{widetext}
\begin{equation}
A_{\pm}=\frac{\zeta\left\{  \left(  p_{\omega}\pm\zeta\right)  \exp (\mp
i\bar{\chi})+\left( p_{\omega}\mp\zeta\right)  \left[  1-r\left(  1-\cos
\eta\right) \right] \right\} -irp_{\omega}\left(  p_{\omega}\mp\zeta\right)
\sin\eta}{2\left( \tilde{\omega}^{2}+i\nu_{c}\tilde{\omega}\right)  \left\{
\left[ p_{\omega}^{2}+\zeta^{2}\right]  \sin\bar{\chi}+2ip_{\omega}\zeta\cos
\bar{\chi}\right\}  },
\end{equation}
where $\bar{\chi}\equiv p_{\omega}L$ and $\eta\equiv p_{\omega}W$. This gives
for the boundary phases which determine the outside radiation
\begin{align}
\theta_{\omega}(0) &  =\frac{i}{\tilde{\omega}^{2}+i\nu_{c}\tilde{\omega}
}+\frac{\zeta\left\{  p_{\omega}\left[  1+\cos\bar{\chi}\right]  -i\zeta\sin
\bar{\chi}\right\}  -rp_{\omega}\left\{  \zeta\left(  1-\cos\eta\right)
+ip_{\omega}\sin\eta\right\}  }{\left(  \tilde{\omega}^{2}+i\nu_{c}
\tilde{\omega}\right)  \left[
(p_{\omega}^{2}+\zeta^{2})\sin\bar{\chi}+2ip_{\omega
}\zeta\cos\bar{\chi}\right]  },\label{StepPh0}\\
\theta_{\omega}(L)\! &  =\!\frac{i\left\{  1-r(1-\cos\eta)\right\}  }
{\tilde{\omega}^{2}+i\nu_{c}\tilde{\omega}}\nonumber\\
&  +\!\frac{\zeta\!\left[  p_{\omega}\!\left(  1+\cos\bar{\chi}\right)
\!-\!i\zeta\sin\bar{\chi}\right]  \!-\!r\left[  \zeta\left(
1\!-\!\cos\eta\right) \!+\!ip_{\omega}\sin\left(  \eta\right)  \right]  \left[
p_{\omega}\cos \bar{\chi}\!-\!i\zeta\sin\bar{\chi}\right]  }{\left(
\tilde{\omega}^{2}+i\nu_{c} \tilde{\omega}\right)  \left\{  \left[
p_{\omega}^{2}+\zeta^{2}\right]
\sin\bar{\chi}+2ip_{\omega}\zeta\cos\bar{\chi}\right\} }.\label{StepPhL}
\end{align}
\end{widetext}
Near the resonance, $\tilde{\omega} L=\pi$, the coefficients $A_{+}$ and $A_{-}$
can be strongly simplified
\[
A_{+}\approx A_{-}\approx\frac{-ir\sin\left[  \tilde{\omega}W\right]  /\left(
2\omega^{2}\right) }{2(\tilde{\omega}-\pi/L)-i(\nu_{c}+\nu_{r})}.
\]
In the resonance $A_{+}\approx A_{-}\approx r\sin\left[  \tilde{\omega}W\right]
/(4\tilde{\omega}^{2}\beta_{\omega})$ giving
\[
\theta_{\omega}(0)\approx\frac{i}{\tilde{\omega}^{2}+i\nu_{c}\tilde{\omega}
}+\frac{r\sin\left[  \tilde{\omega}W\right]  }{2\tilde{\omega}\beta_{\omega}}.
\]
At $W\ll L$ the resonance is strong, $\theta_{\omega}(0)>1$, if $rW\gg2\tilde{\omega}\beta_{\omega}$
or, in real units,
\[
rW\gg\frac{\pi^{2}\lambda_{c}^{2}L_{z}}{\varepsilon_{c}L^{2}}.
\]

The total radiated power in resonance can be estimated as
\begin{equation}
P_{\mathrm{tot}}(\omega_{1})\approx\frac{2L_{y}L^{2}j_{J}^{2}r^{2}\sin^{2}(\pi
W/L)}{\pi\omega},\label{StepRadTot}
\end{equation}
which is consistent with the general formula (\ref{RadPowGenReson}).

The reduced JCC flowing through the stack (\ref{JosCurrGen}) can be computed as
\begin{align}
i_{J}\! &  =\!\frac{\left(  L\!-\!r(2\!-\!r)W\right)  \nu_{c}}{2L\left(
\tilde{\omega}^{2}+\nu_{c}^{2}\right)  \tilde{\omega}}-\frac{\left(
1\!-\!r\right)  r} {2L}\operatorname{Im}\left[
\!\frac{\sin\eta}{p_{\omega}^{3}}\!\right]
\!\nonumber\\
&  +\operatorname{Re}\left[  \!\frac{\mathcal{N}/\bar{\chi}}{\left(
\tilde{\omega}^{2}\!+\!i\nu_{c}\tilde{\omega}\right)  \left\{  \left[
p_{\omega} ^{2}\!+\!\zeta^{2}\right]
\sin\bar{\chi}\!+\!2ip_{\omega}\zeta\cos\bar{\chi}\right\} }\right]
\label{StepJosCurr}
\end{align}
with
\begin{widetext}
\begin{align*}
\mathcal{N} &  =\left(
p_{\omega}\cos\frac{\bar{\chi}}{2}-i\zeta\sin\frac{\bar{\chi}} {2}\right)
\left\{ \zeta\sin\frac{\bar{\chi}}{2}-r\sin\frac{\eta}{2}\left[
\zeta\cos\frac{\eta}{2}+i\left(  p_{\omega}\cos\frac{\eta}{2}-i\zeta\sin
\frac{\eta}{2}\right)  \sin\frac{\bar{\chi}}{2}\right]  \right\}  \\
&  +ir^{2}\sin^{2}\frac{\eta}{2}\left(  p_{\omega}\cos\frac{\eta}{2}
-i\zeta\sin\frac{\eta}{2}\right)  \left(  p_{\omega}\cos\left[  \bar{\chi}
-\frac{\eta}{2}\right]  +i\zeta\sin\left[  \bar{\chi}-\frac{\eta}{2}\right]
\right)
\end{align*}
\end{widetext}

Near the resonance, we estimate $\mathcal{N}\approx-ip_{\omega}^{2}
r^{2}\sin^{2}\left(  \tilde{\omega} W\right)/2$ which gives
\begin{equation}
i_{J}\!\approx\!\frac{r^{2}\sin^{2}\left(  \tilde{\omega} W\right)  \left\{  \nu
_{c}L/2+2\beta_{\omega}\right\}  /2\pi}{\omega^{2}\left\{  \sin^{2}(\tilde{\omega}
L)+\left(  \nu_{c}L/2+2\beta_{\omega}\right)  ^{2}\right\}  }
\label{StepJosCurrRes}
\end{equation}
The maximum current enhancement in the resonance is given by
\[
i_{J,\max}\!\approx\!\frac{r^{2}\sin^{2}\left(  \pi W/L\right)  }{\pi
^{2}\tilde{\omega}\left(  \nu_{c}+4\beta_{\omega}/L\right)  }.
\]
in agreement with Eqs.\ (\ref{JosCurrMax}) and (\ref{StepCouplConst}).

General cumbersome formulas (\ref{StepPh0}), (\ref{StepPhL}), and
(\ref{StepJosCurr}) can be significantly simplified if we assume the
conditions $\nu_{c}$, $|\zeta|\ll\omega$ and $W\ll L$ valid in most practical
situations. In this case these equations can be represented in approximate, simpler form,
\begin{align}
\theta_{\omega}(0)\! &  \approx\frac{i}{\tilde{\omega}^{2}+i\nu_{c}\tilde
{\omega}}-\!\frac{irW/\tilde{\omega}}{\sin\bar{\chi}-i\left(
\nu_{c}L/2\!+\!2\beta
_{\omega}\right)  \cos\bar{\chi}},\\
\theta_{\omega}(L)\! &  \approx\!\frac{i}{\tilde{\omega}^{2}\!+\!i\nu
_{c}\tilde{\omega}}-\!\frac{\!irW\cos\bar{\chi}/\tilde{\omega}}{\sin\bar{\chi}-i\left(
\nu_{c}L/2+2\beta_{\omega}\right)  \cos\bar{\chi}},
\end{align}
and
\begin{widetext}
\begin{align*}
i_{J} &  \approx\nu_{c}\tilde{\omega}+\frac{\nu_{c}}{2\tilde{\omega}^{3}
}-\frac{\nu_{c}rW}{2L\tilde{\omega}^{3}}\\
&  +\operatorname{Re}\left\{ \frac{2\beta_{\omega}\sin\bar{\chi}-i\tilde{\omega
}rW\left[ \sin\bar{\chi}-2i\beta_{\omega}\left[  \cos\left(
\bar{\chi}-\frac{\eta} {2}\right) +\cos\left(  \frac{\eta}{2}\right)  \right]
-2\tilde{\omega }rW\cos\left( \bar{\chi}-\frac{\eta}{2}\right)  \right]
}{2L\tilde{\omega} ^{3}\left[ \sin\bar{\chi}-i\left(
\nu_{c}L/2+2\beta_{\omega}\right)  \cos \bar{\chi}\right] }\right\}
\end{align*}
\end{widetext}
with $\bar{\chi}\approx\tilde{\omega}L$ and$\ \eta\approx\tilde{\omega}W$.

Figure \ref{Fig-StepModWdep} illustrates the evolution of the radiated power
and resonance feature in the I-V dependence with increasing width of the
suppressed region. The JCC density in the suppressed region is assumed to be
half of its value in the rest part, $r=0.5$. For used parameters, the maximum
radiation power in this plot is around, $P_{\mathrm{max}}/L_{y}\sim 0.05$W/cm.

%\begin{figure*}[ptb]
%\begin{center}
%\includegraphics[width=5in]{AsymModulPlots1.png}
%\end{center}
%\caption{The Josephson-frequency (or voltage) dependencies of the current
%density $j$ (in units of the Josephson current density), irradiated power $P$
%(in units of $\Phi_{0}^{2}\omega_{p}^{3}N^{2}/64\pi^{3}c^{2}$), the fraction
%of supplied power converted to radiation $Q$, and the amplitude of oscillating
%phase at the boundary. The following parameters have been used $L=0.6$,
%$W=0.01$L, $r=0.5$ $\nu_{c}=0.002$, and $\mathrm{Re}[\zeta_{N} ]=10^{-4}
%\tilde{\omega}^{2}$ (corresponding to $N\approx600$).}
%\end{figure*}

\section{Discussion and Summary\label{}}

Let us discuss now practical ways to prepare mesas with lateral modulation of
the critical current density. Mesas with linear JCC modulation can be
fabricated in a crystal with inhomogeneous doping. One possible way to prepare
such inhomogeneity is to utilize the sensitivity of doping in BSCCO to the
oxygen concentration. Due to strong temperature dependence of the oxygen
diffusivity \cite{YangAPL99}, in principle, the oxygen concentration profile in
the crystal can be prepared  by short-time annealing by carefully selecting the
annealing temperature and time. In a similar way, mesas with parabolic-like
profiles can be prepared by short-time annealing of mesas themselves already
after fabrication. Another way to prepare modulation in a controlled way is to
use radiation with high-energy electrons, protons, or heavy ions. If part of
the mesa is protected by a mask, this radiation will produce a mesa with
steplike suppression of the critical current.

%Finally, we would like to discuss technical challenges in implementation of the
%proposed device. Clearly,
The major technical challenge is to prepare a mesa with significant modulation
of the Josephson coupling \emph{identical in all junctions}. Variation of
parameters in different junctions, which may be caused by composition
variations, inhomogeneous heating, and different junction areas would strongly
reduce the optimal performance. The quantitative analysis of the radiation
properties of mesas with such parameter variations in different layers will be
done elsewhere.

As the optimal mesa size is rather large, another major technical problem is
sample heating due to quasiparticle damping.  The self-heating in the BSCCO
mesas has been investigated by many experimental groups \cite{Heating}. The
major focus of these studies was the influence of heating on the gap feature in
I-V characteristics, which is located at voltages 30-60 mV/junction. Even
though our voltage range is significantly lower, $\sim$2 mV/junction, the
heating is still expected to be significant due to the required large lateral
size of the mesa. For example, for $\sigma_c = 0.003$ $1/$[$\Omega$~cm],
$N=1000$, and $L_y=300\mu$m, 10 mW of power will be dissipated inside the mesa.
This heat has to be removed from the mesa faces. Therefore, efficient heat
removal is crucial for operation of the device. Recent experimental
observations of the resonant emission using underdoped BSCCO \cite{LutfiSci07}
demonstrate that the heating effects can be manageable even in large-size mesas
with lateral sizes of several hundred micrometers in the voltage range
corresponding to the Josephson frequencies around 1 THz.

The designs with improved thermal management may include, for example,
fabrication of underdoped mesas on the top of overdoped crystal, using massive
gold contacts on the top and bottom of the mesa, and placing an insulator with
high thermal conductivity, such as sapphire, in good thermal contact at the
side of the mesa. From these considerations, mesas with asymmetric modulation
look more preferable than ones with symmetric modulation, because they need a
smaller lateral size for the lowest resonance mode. In the case of symmetric
modulation, the design with suppression of the JCC in the middle, $r<0$, looks
more practical for better thermal management. In fact, the material in the
middle can even be made insulating, because this part is needed only to form
almost standing wave at the working frequency. To excite resonance mode, it is
sufficient to have superconducting regions only at the edges.

In conclusion, we demonstrated that a stack of the intrinsic Josephson
junctions with modulated Josephson coupling represents a very powerful and
efficient source of electromagnetic radiation at the resonance frequency set by
its lateral size. Selecting this size, the generation frequency can be tuned to
the terahertz range. Power levels up to several milliwatts look plausible in
such structures.

\section{Acknowledgements}

AEK would like to acknowledge very useful discussions and joint work on
practical implementation of the device discussed in this manuscript with U.\
Welp, K.\ Gray, L.\ Ozyuzer, and C. Kurter. In Argonne this work was supported
by the Department of Energy under contract No. DE-AC02-06CH11357. In the Los
Alamos National Laboratory this work was carried out under the auspices of the
National Nuclear Security Administration of the Department of Energy under
contract No. DE-AC-06NA25396.

\appendix

\section{Boundary conditions for the homogeneous oscillating phase
\label{App-Bound}}

In this appendix, we consider the boundary conditions for the oscillating phase
at the edges and the radiation power for a stack of intrinsic Josephson
junctions. We will limit ourselves to the case when the oscillating phase is
identical in all junctions. A more general case will be considered elsewhere.
The oscillating phase $\theta_{\omega}$ defined by Eq.\ (\ref{PhaseResSt}) is
connected with the electric and magnetic fields by
the Josephson relations%
\begin{align}
&  E_{z}=-\frac{i\omega \Phi_{0}}{2\pi cs}\theta_{\omega},\label{E-Jos}\\
&  B_{y}=\frac{\Phi_{0}}{2\pi s}\nabla_{x}\theta_{\omega}, \label{h_Jos}%
\end{align}
Therefore, the boundary conditions for the oscillating phase at the edges are
determined by the relation between the fields $E_{z}$ and $B_{y}$ in the
outside media, which we assume to be monochromatic with time dependences
$\propto\exp(-i\omega t)$.

%For illustration, we consider first the simplest geometry when the crystal
%sizes $L_{y}$ and $L_{z}$ along the $y$ and $z$ axes are larger than wavelength
%of radiation in the outside media, $\lambda_{0}$ (the condition
%$L_{z}>\lambda_{0}$ is clearly not realistic for BSCCO mesas).
Outside dielectric media at $|x-L/2|>L/2$ is characterized by the dielectric
constant $\varepsilon_{d}$, and we assume only outgoing wave in this space. The
Fourier components of fields with $|k_{z}|<\sqrt{\varepsilon_{d} }k_{\omega}$
propagate in the media, while the field components with
$|k_{z}|>\sqrt{\varepsilon_{d}}k_{\omega}$ decay. In particular, for
$E_{z}(\omega,x,k_{z})$ at $x>L$, we have
\begin{equation}
E_{z}(\omega,x,k_{z})=E_{z}(\omega,L,k_{z})\exp\left[  ik_{x}(\omega ,k_{z})
(x-L)  \right] \label{Ez-kz}
\end{equation}
with
\begin{equation}
k_{x}(\omega,k_{z})\!=\!%
\genfrac{\{}{.}{0pt}{}{\sqrt{\varepsilon_{d}k_{\omega}^{2}-k_{z}^{2}
}\mathrm{sign}(\omega),\ \text{ for }|k_{z}|\!<\!\sqrt{\varepsilon_{d}%
}|k_{\omega}|,}{i\sqrt{k_{z}^{2}-\varepsilon_{d}k_{\omega}^{2}},\ \text{
for }|k_{z}|\!>\!\sqrt{\varepsilon_{d}}\left\vert k_{\omega}\right\vert .}%
\end{equation}
Other field components, $E_{x}$ and $B_{y}$, are also expressed via
$E_{z}(\omega,L,k_{z})$. First, $E_{x}(\omega,x,k_{z})$ is obtained from Eq.\
(\ref{Ez-kz}) and the Maxwell equation $\nabla\cdot\mathbf{E}=0$, and then
$B_{y}(\omega,x,k_{z})$ is obtained using the Maxwell equation $(\nabla
\times\mathbf{E})_{y}=ik_{\omega}B_{y}$ leading to the following result
\begin{equation}
B_{y}(\omega,\!x,\!k_{z})\!=\!-E_{z}(\omega,\!L,\!k_{z})\frac{\varepsilon_{d}k_{\omega}%
}{k_{x}\!(\omega,\!k_{z})}\exp\left[ik_{x}\!(\omega,\!k_{z})\left(
x\!-\!L\right) \right].
\end{equation}
%\begin{eqnarray}
%&&B_{y}(\omega,x,k_{z}) =\frac{\varepsilon_{d}|k_{\omega}|E_{y}(\omega,L_x/2,k_{z})}{\sqrt
%{\varepsilon_{d}k_{\omega}^{2}-k_{z}^{2}}}\exp
%[i\sqrt{\varepsilon_{d}k_{\omega}^{2}-k_{z}^{2}}\mathrm{sign}(\omega)(x-L_x/2)], \ \ \ \text{
%for }|k_{z}|<\sqrt{\varepsilon_{d}}k_{\omega},\nonumber\\
%&&B_{y}(\omega,x,k_{z})
%=\frac{-i\varepsilon_{d}k_{\omega}E_{z}(\omega,L_x/2,k_{z})}{\sqrt{k_{z}^{2}-\varepsilon_{d}
%k_{\omega}^{2}}}\exp[-\sqrt{k_{z}^{2}-\varepsilon
%_{d}k_{\omega}^{2}}(x-L_x/2)], \ \ \ \text{ for
%}|k_{z}|>\sqrt{\varepsilon_{d}}k_{\omega},\label{Bz_ky}
%\end{eqnarray}
This gives the relation between the fields at the boundary $x=L$,
\begin{align}
&  B_{y}(L,k_{z})=-\zeta(\omega,k_{z})E_{z}(L,k_{z}),\label{ByEz-kz}\\
&  \zeta(\omega,k_{z})=%
\genfrac{\{}{.}{0pt}{}{|k_{\omega}|\varepsilon_{d}/\sqrt{\varepsilon
_{d}k_{\omega}^{2}-k_{z}^{2}}\text{, for }|k_{z}|<\sqrt{\varepsilon_{d}%
}|k_{\omega}|,}{-ik_{\omega}\varepsilon_{d}/\sqrt{k_{z}^{2}-\varepsilon
_{d}k_{\omega}^{2}}\text{, for }|k_{z}|>\sqrt{\varepsilon_{d}}\left\vert
k_{\omega}\right\vert .}%
\nonumber
\end{align}
The condition at $x=0$ has opposite sign, $B_{y}(0,k_{z})=\zeta(\omega
,k_{z})E_{z}(0,k_{z})$. Note again that the term $\zeta(\omega,k_{z})$ for
$|k_{z}|<\sqrt{\varepsilon_{d}}k_{\omega}$ originates from outgoing
electromagnetic wave (radiation), while the term $\zeta(\omega,k_{z})$ for
$|k_{z}|>\sqrt{\varepsilon_{d}}k_{\omega}$ is due to the wave decaying at
distance $\sim(k_{z}^{2}-\varepsilon_{d}k_{\omega}^{2})^{-1/2}$ from the
crystal boundary. The latter term does not carry energy out of the junctions.
In particular, for $k_{z}=0$, we have $B_{y}(L,0)=-\sqrt{\varepsilon_{d}}%
E_{z}(L,0)$, leading to the simple boundary condition for the homogeneous
oscillating phase in the limit $L_{y},L_{z}\gg\lambda_{0}$, $\nabla_{x}%
\theta_{\omega}=\pm(i\sqrt{\varepsilon_{d}}\omega/c)\theta_{\omega}$ for
$x=L,0$. The relation (\ref{ByEz-kz}) can also be rewritten in the
frequency-space representation as
\begin{align}
&B_{y}(L,z,\omega)   =-\int_{-\infty}^{\infty}dz^{\prime}U(z-z^{\prime
},\omega)E_{z}(L,z^{\prime},\omega),\label{ByEz-z}\\
&U(z,\omega)  =-\frac{\varepsilon_{d}}{2}\left[  |k_{\omega}|J_{0}%
(\sqrt{\varepsilon_{d}}|k_{\omega}z|)+ik_{\omega}N_{0}(\sqrt{\varepsilon_{d}%
}|k_{\omega}z|)\right]  ,\nonumber
\end{align}
where $J_{0}(z)$ and $N_{0}(z)$ are the Bessel functions.

The same approach can be used in the realistic case of a crystal small along
the $z$ axis, $L_{z}<\lambda_{0}$, if we know the radiated electric field at
the planes $x=0,L$ outside of the crystal, at $|z|>L_{z}/2$. If we put well
conducting screens there, we can approximate $E_{z}=0$ at $|z|>L_{z}/2$. In
this case for the homogeneous $n$-independent electric field, we obtain for
the average magnetic field at the edge%
\begin{equation}
\overline{B_{y}}(L,\omega)\approx-\frac{L_{z}\varepsilon_{d}}{2}\left[
|k_{\omega}|-\frac{2i}{\pi}k_{\omega}\ln\frac{C}{\sqrt{\varepsilon_{d}}%
L_{z}|k_{\omega}|}\right]  E_{z}(L,\omega),
\end{equation}
with $C=2\exp(3/2-\gamma_{E})\approx5.03$, where $\gamma_{E}\approx0.5772$ is
the Euler constant. We can see that for a small-size mesa, the magnetic field
at the boundary is reduced by the factor $\sim L_{z}k_{\omega}$ in comparison
to the infinite-$L_{z}$ case. This gives the following boundary condition for
the oscillating phase%
\begin{equation}
\nabla_{x}\theta_{\omega}\!=\!\pm\frac{ik_{\omega}L_{z}\varepsilon_{d}}{2}\left[
|k_{\omega}|\!-\!\frac{2i}{\pi}k_{\omega}\ln\frac{C}{\sqrt{\varepsilon_{d}}%
L_{z}|k_{\omega}|}\right]  \theta_{\omega},\
\end{equation}
for $x=L,0$. This corresponds to the boundary conditions (\ref{BoundCond}) and
(\ref{betaN}) in reduced coordinates and $\varepsilon_{d}=1$ used in the paper.
Therefore, for short crystals, the boundary condition can not be written in the
form of an instantaneous relation in between the space and time derivatives of
the phase. This significantly complicates their numerical implementation.
Screens also completely isolate semispaces $x>L$ and $x<0$ and eliminate
interference of radiation coming from the opposite edges.

%In more common situation of a short stack without screens mixing of radiation
%coming from the opposite edges has to be taken into account. In this case the
%relation between the magnetic and electric field at the edges can be written
%as%
%\[
%B_{y}(\alpha L/2,\omega)=\sum_{\beta=\pm1}Z_{\alpha\beta}E_{z}(\beta
%L/2,\omega)
%\]%
%\[
%P_{\mathrm{right}}=-\frac{c}{8\pi}\operatorname{Re}\left[  B_{y}%
%(L/2)E_{z}^{\ast}(L/2)\right]  =-\frac{c}{8\pi}\left\{  \operatorname{Re}%
%\left[  Z_{++}\right]  |E_{z}(L/2,\omega)|^{2}+\operatorname{Re}\left[
%Z_{+-}E_{z}(-L/2,\omega)E_{z}^{\ast}(L/2)\right]  \right\}
%\]
%

\section{Radiated power from a rectangular mesa
\label{App-Rad}}

The radiation power from a short rectangular mesa can be found approximately.
For such a mesa radiation influences weakly the shape of the resonance mode. In
such a situation, the radiation is mostly determined by the distribution of the
oscillating electric field at the mesa edge, which, in turn, is determined by
the shape of the internal mode. Finding the radiation occurs to be a somewhat
easier problem than finding general boundary conditions for the oscillating
phase. An approximate expression for the radiated electric field far away from
the crystal and, thus, the radiation power can be calculated using the
Huyghens' principle, as it is developed in the theory of antennas, see, e.g.,
Ref. \onlinecite{Elliott}. This approach has been applied to resonance modes in
rectangular capacitors in Ref.\ \onlinecite{LeoneIEEE03}. Such a consideration
clearly shows the role of screens and crystal geometry in the formation of the
radiation. The Huyghens' principle in the formulation of Schelkunoff
(equivalence principle) states that we can find fields outside of real sources
(currents and charges) if we know equivalent sources placed on some boundary
surface surrounding real sources. In particular, the equivalent magnetic
current, $\mathbf{M}_{s}$, is related to the tangential components of the
electric field $\mathbf{E}$ on the surface as
\begin{equation}
\mathbf{M}_{s}=-\frac{c}{4\pi}\mathbf{n}\times\mathbf{E}, \label{MagCurr}%
\end{equation}
where $\mathbf{n}$ is the normal vector on the boundary surface.

\begin{figure}[ptb]
\begin{center}
\includegraphics[width=2.5in]
{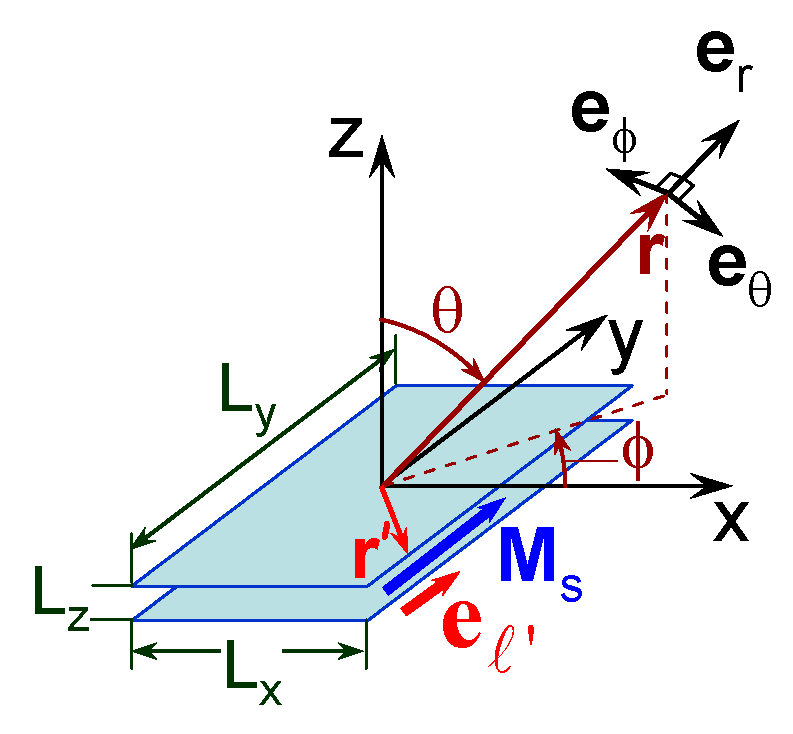}
\end{center}
\caption{(Color online) Geometry for radiation out of a rectangular mesa.}%
\label{Fig-RadRectGeom}%
\end{figure}
In the following, we consider the crystal inside the volume $0<x<L_{x}$,
$0<y<L_{y}$, and $0<z<L_{z}$ bounded by thin metallic contacts on the top and
bottom, see Fig.\ \ref{Fig-RadRectGeom}. The contacts are highly conductive,
and there we can neglect the tangential electric field. At the crystal edges
$x=0,L_{x}$ we can neglect the magnetic field when the radiation power is
small, as in the case of radiation from a capacitor with a small distance
$L_{z}$ between electrodes, $k_{\omega}L_{z}\ll1$, see
Ref.~\onlinecite{LeoneIEEE03}. In this case, we need to account only for the
electric field at the crystal edges, which produces the magnetic equivalent
currents (\ref{MagCurr}). They are related to the oscillating phases
$\theta_{\omega}$ at these edges according to Eq.~(\ref{E-Jos}). We consider
modes inside the crystal which are uniform along the $z$-axis (synchronized
Josephson oscillations in all IJJs). In this approximation, the electric field
$E_{z}$ inside the crystal is described by standing waves characterized by the
indices $m$ and $n$:
\begin{align}
E_{z}(\mathbf{r})&=E_{z}(m,n)\cos(k_{x,m}x)\cos(k_{y,n}y),\label{mode-mn}\\
k_{x,m}&=\pi m/L_{x},\ \ k_{y,n}=\pi n/L_{y} \nonumber
\end{align}
Faraway radiated electric field in terms of $E_{z}(n,m)$ is given by the
expression
\[
\mathbf{E}\!=\!-\frac{k_{\omega}L_{z}}{c}\frac{\exp(ik_{\omega}r)}{r}\int
\!M_{s}(\mathbf{r}^{\prime})\exp(-ik_{\omega}\mathbf{r}^{\prime
}\mathbf{e}_{r})(\mathbf{e}_{r}\times\mathbf{e}_{\ell^{\prime}})d\ell^{\prime
},
\]
where the integral is taken over the perimeter of the crystal, and the
coordinate system as well as definitions of the unit vectors $\mathbf{e}_{r}$
and $\mathbf{e}_{\ell^{\prime}}$ are given in Fig.~\ref{Fig-RadRectGeom}.
Integration over contour $\ell$ gives the following result\cite{LeoneIEEE03}
\begin{align*}
\frac{\mathbf{E}}{E_{z}(m,n)}&=\frac{L_{z}\exp(ik_{\omega}r)}{4\pi r}\\
\times &\left(
\mathbf{P}_{x}\frac{k_{\omega}k_{\xi}}{k_{x,m}^{2}-k_{\xi}^{2}}-\mathbf{P}%
_{y}\frac{k_{\omega}k_{\eta}}{k_{y,n}^{2}-k_{\eta}^{2}}\right)  G_{x}G_{y},
\end{align*}
where
\begin{align*}
\mathbf{P}_{x} &  =\sin\phi\ \mathbf{e}_{\theta}+\cos\theta\cos\phi
\ \mathbf{e}_{\phi},\\
\mathbf{P}_{y} &  =-\cos\phi\ \mathbf{e}_{\theta}+\cos\theta\sin \phi\
\mathbf{e}_{\phi},
\end{align*}
and
\[
\genfrac{\{}{\}}{0pt}{}{k_{\xi}}{k_{\eta}} =k_{\omega}\sin\theta
\genfrac{\{}{\}}{0pt}{}{\cos\phi}{\sin\phi}.
\]
The interference factors $G_{x}=1-(-1)^{n}\exp\left(  ik_{\xi}L_{x}\right)  $
and $G_{y}=1-(-1)^{m}\exp\left(  ik_{\eta}L_{y}\right)  $ describe the
contribution of waves coming to the faraway point $\mathbf{r}$ from opposite
sides of the crystal $x^{\prime}=0$ and $x^{\prime}=L_{x}$ along the $x$ axis
as well as from opposite sides $y^{\prime}=0$ and $y^{\prime}=L_{y}$,
respectively. Their role in the formation of the radiation becomes clear if we
will consider different modes. From the radiated electric field, we can compute
the total radiated power as%
\begin{equation}
P=\frac{c}{8\pi}r^{2}\int_{-\pi}^{\pi}d\phi\int_{0}^{\pi}\sin\theta
d\theta\left[  |E_{\theta}|^{2}+|E_{\phi}|^{2}\right]  .\label{RadPower}%
\end{equation}

For homogeneous oscillations $m=n=0$, we obtain
\begin{widetext}
\[
\frac{\mathbf{E}}{E_{z}(0,0)}=-\frac{L_{z}\exp(ik_{\omega}r)}{4\pi r}\left(
\frac{k_{\omega}}{k_{\xi}}\mathbf{P}_{x}-\frac{k_{\omega}}{k_{\eta}}%
\mathbf{P}_{y}\right)  \left[  1-\exp\left(  -ik_{\xi}L_{x}\right)  \right]
\left[  1-\exp\left(  -ik_{\eta}L_{y}\right)  \right]
\]
corresponding to
\begin{equation}
\frac{E_{\theta}}{E_{z}(0,0)}=-\frac{L_{z}\exp(ik_{\omega}r)}{4\pi r}%
\frac{\left[  1-\exp\left(  -ik_{\omega}L_{x}\sin\theta\cos\phi\right) \right]
\left[  1-\exp\left(  -ik_{\omega}L_{y}\sin\theta\sin\phi\right) \right]
}{\sin\theta\cos\phi\sin\phi},
\end{equation}
and $E_{\phi}=0$. The radiated power (\ref{RadPower}) is given by
\[
P=\frac{c|E_{z}(0,0)|^{2}}{32\pi^{3}}\int_{-\pi}^{\pi}d\phi\int_{0}^{\pi
}d\theta\frac{\left[  1-\cos\left(  k_{\omega}L_{x}\sin\theta\cos\phi\right)
\right]  \left[  1-\cos\left(  k_{\omega}L_{y}\sin\theta\sin\phi\right)
\right]  }{\sin\theta\cos^{2}\phi\sin^{2}\phi}%
\]
\end{widetext}
For small-size crystal $L_{x},L_{y}\ll k_{\omega}^{-1}$ with almost uniform
Josephson oscillations, we obtain
\begin{equation}
P=\frac{c|E_{z}(0,0)|^{2}}{48\pi^{2}}k_{\omega}^{4}L_{x}^{2}L_{y}^{2}L_{z}%
^{2}.
\end{equation}
This is the result for dipole radiation because all sizes of the crystal are
small in comparison with the wavelength of the radiated field. For the crystal
with size $L_{y}$ bigger than the radiation wavelength, $k_{\omega}L_{y}\gg1$,
the result is quite different:
\begin{equation}
P\approx\frac{\omega L_{y}L_{z}^{2}|E_{z}(0,0)|^{2}}{16\pi}\left[
1-J_{0}\left(  k_{\omega}L_{x}\right)  \right]  . \label{RadHomMode}%
\end{equation}
Now the waves coming from opposite sides of the crystal along the $y$ axis do
not interfere with each other and radiation power becomes proportional to
$L_{y}$. For $k_{\omega}L_{x}\ll1$, we obtain the power proportional to
$k_{\omega}^{2}L_{x}^{2}$ due to destructive interference of the waves coming
from opposite sides of the crystal along the $x$ axis. If we put highly
conductive metallic screens separating the spaces $x>L_{x}$ and $x<0$ so that
the edge $x=0$ radiates only into $x<0$ half-space, while that at $x=L_{x}$
radiates only into $x>L_{x}$ half-space (see Fig.~2), the interference will be
eliminated. Such screens also double the radiation coming from one side. This
can be demonstrated in the simplest way using image technique\cite{Elliott}:
radiation from the real electric currents induced at the screens is equivalent
to radiation from the image magnetic current placed next to the original
magnetic current. This leads to doubling of the effective magnetic current,
$M_{s}\rightarrow2M_{s}$, and quadruples the radiated power density. As the
radiation now is limited only by half-space, the total radiated power doubles.
Therefore, in the presence of screens, the factor $[1-J_{0}(k_{\omega}L_{x})]$
in Eq. (\ref{RadHomMode}) has to be replaced by the factor $2$. This means that
the screens strongly enhance the radiation induced by the homogeneous mode in
the case $k_{\omega}L_{x}\ll1$. Such a design with screens for  a crystal thin
along the $x$ axis was proposed in Ref.~\onlinecite{BulKoshPRL07}. This design
gives the possibility of frequency tuning. In addition, heating is reduced due
to small $L_{x}$. However, the crystal should have a large number of layers to
synchronize oscillations in all junctions and work in the super-radiation
regime.

Next, we consider the fundamental cavity mode $(m,n)=(1,0)$, more relevant for
the subject of this paper. In this case, we obtain
\begin{widetext}
\[
\frac{\mathbf{E}}{E_{z}(1,0)}=\frac{L_{z}\exp(ik_{\omega}r)}{4\pi r}\left(
\mathbf{P}_{x}\frac{k_{\omega}k_{\xi}}{(\pi/L_{x})^{2}-k_{\xi}^{2}}%
+\mathbf{P}_{y}\frac{k_{\omega}}{k_{\eta}}\right)  \left[  1-\exp\left(
-ik_{\xi}L_{x}\right)  \right]  \left[  1+\exp\left(  -ik_{\eta}L_{y}\right)
\right]
\]
The components of the faraway electric field are given by the expressions
\begin{align*}
\frac{E_{\theta}}{E_{z}(1,0)} &  =-\frac{L_{z}\exp(ik_{\omega}r)}{4\pi r}%
\frac{\sin^{2}\theta-(\pi/a_{x})^{2}}{\sin^{2}\theta\cos^{2}\phi-(\pi
/a_{x})^{2}}\frac{\cos\phi}{\sin\theta\sin\phi}\left[  1+\exp\left(
-ia_{x}\sin\theta\cos\phi\right)  \right]  \left[  1-\exp\left(  -ia_{y}%
\sin\theta\sin\phi\right)  \right]  ,\\
\frac{E_{\phi}}{E_{z}(1,0)} &  =-\frac{L_{z}\exp(ik_{\omega}r)}{4\pi r}%
\frac{(\pi/a_{x})^{2}}{\sin^{2}\theta\cos^{2}\phi-(\pi/a_{x})^{2}}\frac
{\cos\theta}{\sin\theta}\left[  1+\exp\left(  -ia_{x}\sin\theta\cos \phi\right)
\right]  \left[  1-\exp\left(  -ia_{y}\sin\theta\sin\phi\right) \right]
\end{align*}
with $a_{x}=k_{\omega}L_{x}$ and $a_{y}=k_{\omega}L_{y}$. The radiated power
(\ref{RadPower}) can be represented as
\begin{equation}
P=\frac{cL_{z}^{2}|E_{z}(1,0)|^{2}}{4\pi^{3}}\mathcal{I}_{1,0}(k_{\omega}%
L_{x},k_{\omega}L_{y})\label{Rad10}%
\end{equation}
with%
\begin{align}
\mathcal{I}_{1,0}(a_{x},a_{y}) &  =\int_{0}^{\pi/2}d\phi\int_{0}^{\pi
/2}d\theta\frac{\left[  1+\cos\left(  a_{x}\sin\theta\cos\phi\right)  \right]
\left[  1-\cos\left(  a_{y}\sin\theta\sin\phi\right)  \right]  }{\sin
\theta\sin^{2}\phi}\nonumber\\
&  \times\frac{\left(  \sin^{2}\theta-(\pi/a_{x})^{2}\right)  ^{2}\cos^{2}%
\phi+(\pi/a_{x})^{4}\cos^{2}\theta\sin^{2}\phi}{\left(  \sin^{2}\theta\cos
^{2}\phi-(\pi/a_{x})^{2}\right)  ^{2}}.\label{I10}%
\end{align}
\end{widetext}
In the regime $k_{\omega}L_{y}\gg1$ this gives the following result
\begin{equation}
P\approx\frac{\omega L_{y}L_{z}^{2}|E_{z}(1,0)|^{2}}{16\pi}\left[
1+J_{0}(k_{\omega}L_{x})\right].\label{RadFundMode}%
\end{equation}
Now we have a constructive interference of waves coming from opposite sides of
the crystal along the $x$ axis because electric field on these sides has
opposite signs generating the same-sign magnetic fields. For such mode, screens
do not influence much the radiation in the limit $k_{\omega}L_{x}\ll1$.\ The
reduced parameter of radiation damping $\nu_{r}$ introduced in Eqs.
(\ref{ModeSolutions}) and (\ref{RadDampPar}) is related to the radiation power
as
\begin{equation}
P=\nu_{r}L_{x}L_{y}L_{z}\frac{\varepsilon_{c}\omega_{p}}{16\pi}|E_{z}
(1,0)|^{2}.\label{RadDPar-P}
\end{equation}

The above results can also be straightforwardly generalized to the case when a
stack is bounded by large-size ground plate at the bottom, $z=0$. This is the
case, for example, for the mesa fabricated on the top of bulk crystal. If we
treat the ground plate as an ideal conductor, its influence can again be taken
into account by the image technique\cite{Elliott}. This just leads to the
doubling of the effective magnetic current, $\mathbf{M}_{s}\rightarrow
2\mathbf{M}_{s}$, and to the doubling of the total radiated power
$P\rightarrow2P$.


\begin{thebibliography}{99}
\bibitem{ftnt-underdp}Throughout the paper, in estimates we use parameters
typical for moderately underdoped BSCCO at low temperatures. Usage of the
underdoped BSCCO in the discussed design is preferable for reducing heating due
to smaller quasiparticle conductivity.
\bibitem{RadDampNote}Note that, in general, the radiation damping parameter $\nu_r$
depends not only on stack geometry but also on the excited mode. Only in the
case considered here ($k_{\omega}L_y\gg 1$ and no interference between
radiation from the opposite edges) $\nu_r$ is the same for all $m$.
\end{thebibliography}
\end{document}